# Quantum Bridge Analytics I: A Tutorial on Formulating and Using QUBO Models

**Fred Glover[1], Gary Kochenberger[2], Yu Du[2]**

## Abstract

Quantum Bridge Analytics relates generally to methods and systems for hybrid classical-quantum computing, and more particularly is devoted to developing tools for bridging classical and quantum computing to gain the benefits of their alliance in the present and enable enhanced practical application of quantum computing in the future.

This is the first of a two-part tutorial that surveys key elements of Quantum Bridge Analytics and its applications, with an emphasis on supplementing models with numerical illustrations. In Part 1 (the present paper) we focus on the Quadratic Unconstrained Binary Optimization (QUBO) model which is presently the most widely applied optimization model in the quantum computing area, and which unifies a rich variety of combinatorial optimization problems.

-------------------------------------------------------------------------------------------------------------------

[1]ECEE, College of Engineering and Applied Science, University of Colorado, Boulder, CO 80302 USA  fred.glover@colorado.edu

[2]College of Business, University of Colorado at Denver, Denver, CO 80217 USA, gary.kochenberger@ucdenver.edu; yu.du@ucdenver.edu

## Section 1:  Introduction

The field of Combinatorial Optimization (CO) is one of the most important areas in the field of optimization, with practical applications found in every industry, including both the private and public sectors.  It is also one of the most active research areas pursued by the research communities of Operations Research, Computer Science and Analytics as they work to design and test new methods for solving real world CO problems.

Generally, these problems are concerned with making wise choices in settings where a large number of yes/no decisions must be made and each set of decisions yields a corresponding objective function value – like a cost or profit value.  Finding good solutions in these settings is extremely difficult.  The traditional approach is for the analyst to develop a solution algorithm that is tailored to the mathematical structure of the problem at hand.  While this approach has produced good results in certain problem settings, it has the disadvantage that the diversity of applications arising in practice requires the creation of a diversity of solution techniques, each with limited application outside their original intended use.

In recent years, we have discovered that a mathematical formulation known as QUBO, an acronym for a Quadratic Unconstrained Binary Optimization problem, can embrace an exceptional variety of important CO problems found in industry, science and government, as documented in studies such as Kochenberger, et. al. (2014) and Anthony, et. al. (2017). Through special reformulation techniques that are easy to apply, the power of QUBO solvers can be used to efficiently solve many important problems once they are put into the QUBO framework.

The QUBO model has emerged as an underpinning of the quantum computing area known as quantum annealing and Fujitsu's digital annealing, and has become a subject of study in neuromorphic computing. Through these connections, QUBO models lie at the heart of experimentation carried out with quantum computers developed by D-Wave Systems and neuromorphic computers developed by IBM. The consequences of these new discoveries linking QUBO models to quantum computing are being explored in initiatives by organizations such as IBM, Google, Amazon, Microsoft, D-Wave and Lockheed Martin in the commercial realm and Los Alamos National Laboratory, Oak Ridge National Laboratory, Lawrence Livermore National Laboratory and NASA's Ames Research Center in the public sector. Computational experience is being amassed by both the classical and the quantum computing communities that highlights not only the potential of the QUBO model but also its effectiveness as an alternative to traditional modeling and solution methodologies.

The connection with Quantum Bridge Analytics derives from the gains to be achieved by building on these developments to bridge the gap between classical and quantum computational methods and technologies. As emphasized in the 2019 Consensus Study Report titled Quantum Computing: Progress and Prospects, by the National Academies of Sciences, Engineering and Medicine (https://www.nap.edu/catalog/25196/quantum-computing-progress-and-prospects) quantum computing will remain in its infancy for perhaps another decade, and in the interim

"formulating an R&D program with the aim of developing commercial applications for near-term quantum computing is critical to the health of the field." The report further notes that such a program will rest on developing "hybrid classical-quantum techniques." Innovations that underlie and enable these hybrid classical-quantum techniques are the focus of Quantum Bridge Analytics and draw heavily on the QUBO model for their inspiration.

The significance of the ability of the QUBO model to encompass many models in combinatorial optimization is enhanced by the fact that the QUBO model can be shown to be equivalent to the Ising model that plays a prominent role in physics, as highlighted in in the paper by Lucas (2014). Consequently, the broad range of optimization problems solved effectively by state-of-the-art QUBO solution methods are joined by an important domain of problems arising in physics applications.

The materials provided in the sections that follow illustrate the process of reformulating important optimization problems as QUBO models through a series of explicit examples. Collectively these examples highlight the application breadth of the QUBO model. We disclose the unexpected advantages of modeling a wide range of problems in a form that differs from the linear models classically adopted in the optimization community. We show how many different types of constraining relationships arising in practice can be embodied within the "unconstrained" QUBO formulation in a very natural manner using penalty functions, yielding exact model representations in contrast to the approximate representations produced by customary uses of penalty functions. Each step of generating such models is illustrated in detail by simple numerical examples, to highlight the convenience of using QUBO models in numerous settings. As part of this, we provide techniques that can be used to recast a variety of problems that may not seem at first to fit within an unconstrained binary optimization structure into an equivalent QUBO model. We also describe recent innovations for solving QUBO models that offer a fertile avenue for integrating classical and quantum computing and for applying these models in machine learning.

As pointed out in Kochenberger and Glover (2006), the QUBO model encompasses the following important optimization problems:

- Quadratic Assignment Problems
- Capital Budgeting Problems
- Multiple Knapsack Problems
- Task Allocation Problems (distributed computer systems)
- Maximum Diversity Problems
- P-Median Problems
- Asymmetric Assignment Problems
- Symmetric Assignment Problems
- Side Constrained Assignment Problems
- Quadratic Knapsack Problems

- Constraint Satisfaction Problems (CSPs)
- Discrete Tomography Problems
- Set Partitioning Problems
- Set Packing Problems
- Warehouse Location Problems
- Maximum Clique Problems
- Maximum Independent Set Problems
- Maximum Cut Problems
- Graph Coloring Problems
- Number Partitioning Problems
- Linear Ordering Problems
- Clique Partitioning Problems
- SAT problems

Details of such applications are elaborated more fully in Kochenberger et al. (2014).

In the following development we describe approaches that make it possible to model these and many other types of problems in the QUBO framework and provide information about recent developments linking QUBO to machine learning and quantum computing.

**Basic QUBO Problem Formulation**

We now give a formal definition of the QUBO model whose significance will be made clearer by numerical examples that give a sense of the diverse array of practical QUBO applications. Definition:  The QUBO model is expressed by the optimization problem:

$$QUBO: \text{ minimize/maximize } y = x^t Q x$$

where x is a vector of binary decision variables and Q is a square matrix of constants.

It is common to assume that the Q matrix is symmetric or in upper triangular form, which can be achieved without loss of generality simply as follows:

Symmetric form: For all i and j except i = j, replace $q_{ij}$ by $(q_{ij} + q_{ji})/2$ .

Upper triangular form: For all i and j with $j > i$ , replace  $q_{ij}$ by $q_{ij} + q_{ji}$. Then replace all $q_{ij}$ for $j < i$ by 0. (If the matrix is already symmetric, this just doubles the  $q_{ij}$ values above the main diagonal, and then sets all values below the main diagonal to 0).

In the examples given in the following sections, we will work with the full, symmetric Q matrix rather than adopting the "upper triangular form."

*Comment on the formal classification of QUBO models and their solution:* QUBO models belong to a class of problems known to be NP-hard. The practical meaning of this is that exact solvers designed to find "optimal" solutions (like the commercial CPLEX and Gurobi solvers) will most likely be unsuccessful except for very small problem instances. Using such methods, realistic sized problems can run for days and even weeks without producing high quality solutions. Fortunately, as we disclose in the sections that follow, impressive successes are being achieved by using modern metaheuristic methods that are designed to find high quality but not necessarily optimal solutions in a modest amount of computer time. These approaches are opening valuable possibilities for joining classical and quantum computing.

## Section 2: Illustrative Examples and Definitions

Before presenting common practical applications, we first give examples and definitions to lay the groundwork to see better how these applications can be cast in QUBO form.
To begin, consider the optimization problem

$$\text{Minimize } y = -5x_1 - 3x_2 - 8x_3 - 6x_4 + 4x_1x_2 + 8x_1x_3 + 2x_2x_3 + 10x_3x_4$$

where the variables, $x_j$, are binary. We can make several observations:

1.  The function to be minimized is a quadratic function in binary variables with a linear part $-5x_1 - 3x_2 - 8x_3 - 6x_4$ and a quadratic part $4x_1x_2 + 8x_1x_3 + 2x_2x_3 + 10x_3x_4$.

2.  Since binary variables satisfy $x_j = x_j^2$, the linear part can be written as

$$-5x_1^2 - 3x_2^2 - 8x_3^2 - 6x_4^2$$

3.  Then we can re-write the model in the following matrix form:

$$\text{Minimize } y = \begin{pmatrix} x_1 & x_2 & x_3 & x_4 \end{pmatrix} \begin{bmatrix} -5 & 2 & 4 & 0 \\ 2 & -3 & 1 & 0 \\ 4 & 1 & -8 & 5 \\ 0 & 0 & 5 & -6 \end{bmatrix} \begin{bmatrix} x_1 \\ x_2 \\ x_3 \\ x_4 \end{bmatrix}$$

4.  In turn, this can be written in the matrix notation introduced in Section 1 as

$$\text{Minimize } y = x^t Q x$$

where x is a column vector of binary variables. Note that the coefficients of the original linear terms appear on the main diagonal of the Q matrix. In this case Q is symmetric about the main diagonal without needing to modify the coefficients by the approach shown in Section 1.

5. Other than the 0/1 restrictions on the decision variables, QUBO is an unconstrained model with all problem data being contained in the Q matrix. These characteristics make the QUBO model particularly attractive as a modeling framework for combinatorial optimization problems, offering a novel alternative to classically constrained representations.

6. The solution to the model in (3) above is: $y = -11,\ x_1 = x_4 = 1,\ x_2 = x_3 = 0.$

**Remarks**:

- As already noted, the stipulation that Q is symmetric about the main diagonal does not limit the generality of the model.

- As previously emphasized, a variety of optimization problems can naturally be formulated and solved as an instance of the QUBO model. In addition, many other problems that don't appear to be related to QUBO problems can be re-formulated as a QUBO model. We illustrate this special feature of the QUBO model in the sections that follow.

## Section 3: Natural QUBO Formulations

As mentioned earlier, several important problems fall naturally into the QUBO class. To illustrate such cases, we provide two examples of important applications whose formulations naturally take the form of a QUBO model.

### 3.1 The Number Partitioning Problem

The Number Partitioning problem has numerous applications cited in the Bibliography section of these notes. A common version of this problem involves partitioning a set of numbers into two subsets such that the subset sums are as close to each other as possible. We model this problem as a QUBO instance as follows:

Consider a set of numbers $S = \{s_1, s_2, \dots, s_m\}$. Let $x_j = 1$ if $S_j$ is assigned to subset 1; 0 otherwise. Then the sum for subset 1 is given by $sum_1 = \sum_{j=1}^{m} s_j x_j$ and the sum for subset 2 is given by $sum_2 = \sum_{j=1}^{m} s_j - \sum_{j=1}^{m} s_j x_j$. The difference in the sums is then

$$diff = \sum_{j=1}^{m} s_j - 2\sum_{j=1}^{m} s_j\, x_j = c - 2\sum_{j=1}^{m} s_j\, x_j.$$

We approach the goal of minimizing this difference by minimizing

$$diff^2 = \left\{ c - 2\sum_{j=1}^{m} s_j x_j \right\}^2 = c^2 + 4x^t Q x$$

where

$$q_{ii} = s_i\left(s_i - c\right) \qquad q_{ij} = q_{ji} = s_i s_j$$

Dropping the additive and multiplicative constants, our QUBO optimization problem becomes:

$$QUBO : \min\; y = x^t Q x$$

where the Q matrix is constructed with $q_{ii}$ and $q_{ij}$ as defined above.

**Numerical Example:** Consider the set of eight numbers

$$S = \{25,\, 7, 13,\, 31,\, 42, 17,\, 21, 10\}$$

By the development above, we have $c^2 = 27{,}556$ and the equivalent QUBO problem is $\min y = x^t Q x$ with

$$Q = \begin{bmatrix}
-3525 & 175 & 325 & 775 & 1050 & 425 & 525 & 250 \\
175 & -1113 & 91 & 217 & 294 & 119 & 147 & 70 \\
325 & 91 & -1989 & 403 & 546 & 221 & 273 & 130 \\
775 & 217 & 403 & -4185 & 1302 & 527 & 651 & 310 \\
1050 & 294 & 546 & 1302 & -5208 & 714 & 882 & 420 \\
425 & 119 & 221 & 527 & 714 & -2533 & 357 & 170 \\
525 & 147 & 273 & 651 & 882 & 357 & -3045 & 210 \\
250 & 70 & 130 & 310 & 420 & 170 & 210 & -1560
\end{bmatrix}$$

Solving QUBO gives $x = (0,0,0,1,1,0,0,1)$ for which $y = -6889$, yielding perfectly matched sums which equal 83. The development employed here can be expanded to address other forms of the number partitioning problem, including problems where the numbers must be partitioned into three or more subsets, as discussed in Alidaee, et.al. (2005).

**3.2 The Max-Cut Problem**

The Max Cut problem is one of the most famous problems in combinatorial optimization. Given an undirected graph G(V, E) with a vertex set V and an edge set E, the Max Cut problem seeks to partition V into two sets such that the number of edges between the two sets (considered to be severed by the cut), is a large as possible.

We can model this problem by introducing binary variables satisfying $x_j = 1$ if vertex j is in one set and $x_j = 0$ if it is in the other set. Viewing a cut as severing edges joining two sets, to leave endpoints of the edges in different vertex sets, the quantity $x_i + x_j - 2x_i x_j$ identifies whether the edge $(i, j)$ is in the cut. That is, if $(x_i + x_j - 2x_i x_j)$ is equal to 1, then exactly one of $x_i$ and $x_j$ equals 1, which implies edge $(i, j)$ is in the cut. Otherwise $(x_i + x_j - 2x_i x_j)$ is equal to zero and the edge is not in the cut.

Thus, the problem of maximizing the number of edges in the cut can be formulated as

$$\text{Maximize } y = \sum_{(i,j) \in E} \left( x_i + x_j - 2x_i x_j \right)$$

which is an instance of

$$QUBO : \max y = x^t Q x$$

The linear terms determine the elements on the main diagonal of Q and the quadratic terms determine the off-diagonal elements. See Boros and Hammer (1991, 2002) and Kochenberger et.al.(2013) for further discussions of QUBO and the Max Cut problem.

**Numerical Example:** To illustrate the Max Cut problem, consider the following undirected graph with 5 vertices and 6 edges.

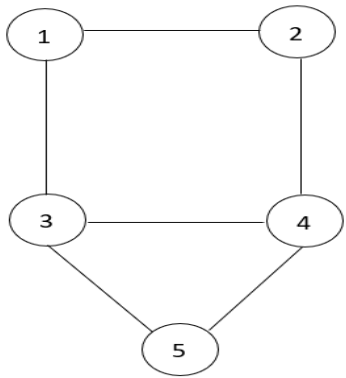

Explicitly taking into account all edges in the graph gives the following formulation:

$$\begin{aligned} Maximize \quad y &= (x_1 + x_2 - 2x_1 x_2) + (x_1 + x_3 - 2x_1 x_3) + (x_2 + x_4 - 2x_2 x_4) \\ &+ (x_3 + x_4 - 2x_3 x_4) + (x_3 + x_5 - 2x_3 x_5) + (x_4 + x_5 - 2x_4 x_5) \end{aligned}$$

or

$$\max\ y = 2x_1 + 2x_2 + 3x_3 + 3x_4 + 2x_5 - 2x_1x_2 - 2x_1x_3 - 2x_2x_4 - 2x_3x_4 - 2x_3x_5 - 2x_4x_5$$

This takes the desired form

$$QUBO: \max\ y = x^t Q x$$

by writing the symmetric Q matrix as:

$$Q = \begin{bmatrix} 2 & -1 & -1 & 0 & 0 \\ -1 & 2 & 0 & -1 & 0 \\ -1 & 0 & 3 & -1 & -1 \\ 0 & -1 & -1 & 3 & -1 \\ 0 & 0 & -1 & -1 & 2 \end{bmatrix}$$

Solving this QUBO model gives $x = (0,1,1,0,0)$. Hence vertices 2 and 3 are in one set and vertices 1, 4, and 5 are in the other, with a maximum cut value of 5.

In the above examples, the problem characteristics led directly to an optimization problem in QUBO form. As previously remarked, many other problems require "re-casting" to create the desired QUBO form. We introduce a widely-used form of such re-casting in the next section.

## Section 4: Creating QUBO Models Using Known Penalties

The "natural form" of a QUBO model illustrated thus far contains no constraints other than those requiring the variables to be binary. However, by far the largest number of problems of interest include additional constraints that must be satisfied as the optimizer searches for good solutions. Many of these constrained models can be effectively re-formulated as a QUBO model by introducing quadratic penalties into the objective function as an alternative to explicitly imposing constraints in the classical sense. The penalties introduced are chosen so that the influence of the original constraints on the solution process can alternatively be achieved by the natural functioning of the optimizer as it looks for solutions that avoid incurring the penalties. That is, the penalties are formulated so that they equal zero for feasible solutions and equal some positive penalty amount for infeasible solutions. For a minimization problem, these penalties are added to create an augmented objective function to be minimized. If the penalty terms can be driven to zero, the augmented objective function becomes the original function to be minimized.

For certain types of constraints, quadratic penalties useful for creating QUBO models are known in advance and readily available to be used in transforming a given constrained problem into a QUBO model. Examples of such penalties for some commonly encountered constraints are given in the table below. Note that in the table, all variables are intended to be binary and the parameter P is a positive, scalar penalty value. This value must be chosen sufficiently large to assure the penalty term is indeed equivalent to the classical constraint, but in practice an acceptable value for P is usually easy to specify. We discuss this matter more thoroughly later.

| Classical Constraint | Equivalent Penalty |
|---|---|
| $x + y \leq 1$ | $P(xy)$ |
| $x + y \geq 1$ | $P(1 - x - y + xy)$ |
| $x + y = 1$ | $P(1 - x - y + 2xy)$ |
| $x \leq y$ | $P(x - xy)$ |
| $x_1 + x_2 + x_3 \leq 1$ | $P(x_1 x_2 + x_1 x_3 + x_2 x_3)$ |
| $x = y$ | $P(x + y - 2xy)$ |

Table of a few Known constraint/penalty pairs

To illustrate the main idea, consider a traditionally constrained problem of the form:

$$Min \; y = f(x)$$
subject to the constraint
$$x_1 + x_2 \leq 1$$

Where $x_1$ and $x_2$ are binary variables. Note that this constraint allows either or neither x variable to be chosen. It explicitly precludes both from being chosen (i.e., both cannot be set to 1).

From the 1st row in the table above, we see that a quadratic penalty that corresponds to our constraint is

$$Px_1 x_2$$

where P is a positive scalar. For P chosen sufficiently large, the unconstrained problem

$$minimize \; y = f(x) + Px_1 x_2$$

has the same optimal solution as the original constrained problem. If f(x) is linear or quadratic, then this unconstrained model will be in the form of a QUBO model. In our present example, any optimizer trying to minimize $y$ will tend to avoid solutions having both $x_1$ and $x_2$ equal to 1, else a large positive amount will be added to the objective function. That is, the objective function incurs a penalty corresponding to infeasible solutions. This simple penalty has been used effectively by Pardalos and Xue (1999) in the context of the maximum clique and related problems.

## 4.1 The Minimum Vertex Cover  (MVC) Problem

In section 3.2 we saw how the QUBO model could be used to represent the famous Max Cut problem.  Here we consider another well-known optimization problem on graphs called the Minimum Vertex Cover problem.  Given an undirected graph with a vertex set V and an edge set E,  a <u>vertex cover</u> is a subset of the vertices (nodes) such that each edge in the graph is incident to at least one vertex in the subset. The Minimum Vertex Cover problem seeks to find a cover with a minimum number of vertices in the subset.

A standard optimization model for MVC can be formulated as follows.  Let $x_j = 1$ if vertex j is in the cover (i.e., in the subset) and $x_j = 0$ otherwise.  Then the standard constrained, linear 0/1 optimization model for this problem is:

$$\text{Minimize } \sum_{j \in V} x_j$$

subject to

$$x_i + x_j \geq 1 \text{ for all } (i, j) \in E$$

Note the constraints ensure that at least one of the endpoints of each edge will be in the cover and the objective function seeks to find the cover using the least number of vertices. Note also that we have a constraint for each edge in the graph, meaning that even for modest sized graphs we can have many constraints.  Each constraint will alternatively be imposed by adding a penalty to the objective function in the equivalent QUBO model.

Referring to our table above, we see that the constraints in the standard MVC model can be represented by a penalty of <u>the form</u> $P(1 - x - y + xy)$. Thus, an unconstrained alternative to the constrained model for MVC is

$$\text{Minimize } y = \sum_{j \in V} x_j + P(\sum_{(i,j) \in E} \left(1 - x_i - x_j + x_i x_j\right))$$

where P again represents a positive scalar penalty. In turn, we can write this as minimize $x^t Q x$ plus a constant term.  Dropping the additive constant, which has no impact on the optimization, we have an optimization problem in the form of a QUBO model.

**Remark**: A common extension of this problem allows a weight $w_j$ to be associated with each vertex j.  Following the development above, the QUBO model for the Weighted Vertex Cover problem is given by:

$$\text{Minimize } y = \sum_{j \in V} w_j x_j + P\left(\sum_{(i,j) \in E} \left(1 - x_i - x_j + x_i x_j\right)\right)$$

**Numerical Example**: Consider the graph of section 3.2 again but this time we want to determine a minimum vertex cover.

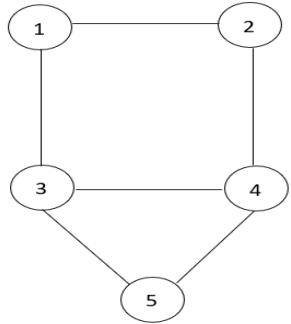

For this graph with n = 6 edges and m = 5 nodes, the model becomes:

$$\begin{aligned}
\textit{Minimize } \quad y &= x_1 + x_2 + x_3 + x_4 + x_5 + \\
&\phantom{=} P(1 - x_1 - x_2 + x_1 x_2) + \\
&\phantom{=} P(1 - x_1 - x_3 + x_1 x_3) + \\
&\phantom{=} P(1 - x_2 - x_4 + x_2 x_4) + \\
&\phantom{=} P(1 - x_3 - x_4 + x_3 x_4) + \\
&\phantom{=} P(1 - x_3 - x_5 + x_3 x_5) + \\
&\phantom{=} P(1 - x_4 - x_5 + x_4 x_5)
\end{aligned}$$

which can be written as

$$\begin{aligned}
\textit{Minimize } \quad y &= (1-2P)x_1 + (1-2P)x_2 + (1-3P)x_3 + (1-3P)x_4 + (1-2P)x_5 \\
&\phantom{=} + Px_1 x_2 + Px_1 x_3 + Px_2 x_4 + Px_3 x_4 + Px_3 x_5 + Px_4 x_5 + 6P
\end{aligned}$$

Arbitrarily choosing P to be equal to 8 and dropping the additive constant (6P = 48) gives our QUBO model

$$QUBO : \min x^t Q x$$

with the Q matrix given by

$$\begin{bmatrix}
-15 & 4 & 4 & 0 & 0 \\
4 & -15 & 0 & 4 & 0 \\
4 & 0 & -23 & 4 & 4 \\
0 & 4 & 4 & -23 & 4 \\
0 & 0 & 4 & 4 & -15
\end{bmatrix}$$

Note that we went from a constrained model with 5 variables and 6 constraints to an unconstrained QUBO model in the same 5 variables. Solving this QUBO model gives: $x^t Q x = -45$ at $x = (0,1,1,0,1)$ for which $y = 48 - 45 = 3$, meaning that a minimum cover is given by nodes 2, 3, and 5. It's easy to check that at this solution, all the penalty functions are equal to 0.

Comment on the Scalar Penalty P:

As we have indicated, the reformulation process for many problems requires the introduction of a scalar penalty P for which a numerical value must be given. These penalties are not unique, meaning that many different values can be successfully employed. For a particular problem, a workable value is typically set based on domain knowledge and on what needs to be accomplished. Often, we use the same penalty for all constraints but there is nothing wrong with having different penalties for different constraints if there is a good reason to differentially treat various constraints. If a constraint must absolutely be satisfied, i.e., a "hard" constraint, then P must be large enough to preclude a violation. Some constraints, however, are "soft", meaning that it is desirable to satisfy them but slight violations can be tolerated. For such cases, a more moderate penalty value will suffice.

A penalty value that is too large can impede the solution process as the penalty terms overwhelm the original objective function information, making it difficult to distinguish the quality of one solution from another. On the other hand, a penalty value that is too small jeopardizes the search for feasible solutions. Generally, there is a 'Goldilocks region' of considerable size that contains penalty values that work well. A little preliminary thought about the model can yield a ballpark estimate of the underlined original objective function value. Taking P to be some percentage (75% to 150%) of this estimate is often a good place to start. In the end, solutions generated can always be checked for feasibility, leading to changes in penalties and further rounds of the solution process as needed to zero in on an acceptable solution.

**4.2 The Set Packing Problem**

The Set Packing problem is a well-known optimization problem in binary variables with a general (traditional) formulation given by

$$\max \sum_{j=1}^{n} w_j x_j$$
$$st$$
$$\sum_{j=1}^{n} a_{ij} x_j \leq 1 \quad for\ i = 1, \ldots m$$

where the $a_{ij}$ are 0/1 coefficients, the $w_j$ are weights and the $x_j$ variables are binary. Using the penalties of the form shown in the first and fifth rows of the table given earlier, we can easily construct a quadratic penalty corresponding to each of the constraints in the traditional model.

Then by subtracting the penalties from the objective function, we have an unconstrained representation of the problem in the form of a QUBO model.

Numerical Example: Consider the following small example of a set packing problem:

$$\max \; x_1 + x_2 + x_3 + x_4$$

$$\text{st}$$

$$x_1 \quad\;\; + x_3 + x_4 \leq 1$$
$$x_1 + x_2 \qquad\quad \leq 1$$

Here all the objective function coefficients, the $w_j$ values, are equal to 1. Using the penalties mentioned above, the equivalent unconstrained problem is:

$$\max \; x_1 + x_2 + x_3 + x_4 - P x_1 x_3 - P x_1 x_4 - P x_3 x_4 - P x_1 x_2$$

This has our customary QUBO form

$$QUBO : \max x^t Q x$$

where the Q matrix , with P arbitrarily chosen to be 6, is given by

$$\begin{bmatrix} 1 & -3 & -3 & -3 \\ -3 & 1 & 0 & 0 \\ -3 & 0 & 1 & -3 \\ -3 & 0 & -3 & 1 \end{bmatrix}$$

Solving the QUBO model gives $y = 2$ at $x = (0,1,1,0)$. Note that at this solution, all four penalty terms are equal to zero.

**Remark**:  Set packing problems with thousands of variables and constraints have been efficiently reformulated and solved in Alidaee, et. al. (2008) using the QUBO reformulation illustrated in this example.

### 4.3 The Max 2-Sat Problem

Satisfiability problems, in their various guises, have applications in many different settings. Often these problems are represented in terms of clauses, in conjunctive normal form, consisting of several true/false literals. The challenge is to determine the literals so that as many clauses as possible are satisfied.

For our optimization approach, we'll represent the literals as 0/1 values and formulate models that can be re-cast into the QUBO framework and solved with QUBO solvers. To illustrate the approach, we consider the category of satisfiability problems known as Max 2-Sat problems. For Max 2-Sat, each clause consists of two literals and a clause is satisfied if either or both literals are true. There are <u>three</u> possible types of clauses for this problem, each with a traditional constraint that must be satisfied if the clause is to be true. In turn, each of these three constraints has a known quadratic penalty given in our previous table.

The <u>three clause types</u> along with their traditional constraints and associated penalties are:

1. <u>No negations:</u> Example $(x_i \lor x_j)$

   Traditional constraint: $x_i + x_j \geq 1$

   Quadratic Penalty: $(1 - x_i - x_j + x_i x_j)$

2. <u>One negation:</u> Example $(x_i \lor \bar{x}_j)$

   Traditional constraint: $x_i + \bar{x}_j \geq 1$

   Quadratic Penalty: $(x_j - x_i x_j)$

3. <u>Two negations:</u> Example $(\bar{x}_i \lor \bar{x}_j)$

   Traditional constraint: $\bar{x}_i + \bar{x}_j \geq 1$

   Quadratic Penalty: $(x_i x_j)$

(Note that $x_j = 1$ or 0 denoting whether literal j is true or false. The notation $\bar{x}_j$, the complement of $x_j$, is equal to $(1 - x_j)$. )

For each clause type, if the traditional constraint is satisfied, the corresponding penalty is equal to zero, while if the traditional constraint is not satisfied, the quadratic penalty is equal to 1. Given this one-to-one correspondence, we can approach the problem of maximizing the number of clauses satisfied by equivalently minimizing the number of clauses not satisfied. This perspective, as we will see, gives us a QUBO model.

For a given Max 2-Sat instance then, we can add the quadratic penalties associated with the problem clauses to get a composite penalty function which we want to minimize. Since the penalties are all quadratic, this penalty function takes the form of a QUBO model, min $y = x^t Q x$. Moreover, if $y$ turns out to be equal to zero when minimizing the QUBO model, this means we have a solution that satisfies all of the clauses; if $y$ turns out to equal 5, that means we have a solution that satisfies all but 5 of the clauses; and so forth. This modeling and solution procedure is illustrated by the following example with 4 variables and 12 clauses where the penalties are determined by the clause type.

| Clause # | Clause | Quadratic Penalty |
|----------|--------|-------------------|
| 1 | $x_1 \vee x_2$ | $(1 - x_1 - x_2 + x_1 x_2)$ |
| 2 | $x_1 \vee \bar{x}_2$ | $(x_2 - x_1 x_2)$ |
| 3 | $\bar{x}_1 \vee x_2$ | $(x_1 - x_1 x_2)$ |
| 4 | $\bar{x}_1 \vee \bar{x}_2$ | $(x_1 x_2)$ |
| 5 | $\bar{x}_1 \vee x_3$ | $(x_1 - x_1 x_3)$ |
| 6 | $\bar{x}_1 \vee \bar{x}_3$ | $(x_1 x_3)$ |
| 7 | $x_2 \vee \bar{x}_3$ | $(x_3 - x_2 x_3)$ |
| 8 | $x_2 \vee x_4$ | $(1 - x_2 - x_4 + x_2 x_4)$ |
| 9 | $\bar{x}_2 \vee x_3$ | $(x_2 - x_2 x_3)$ |
| 10 | $\bar{x}_2 \vee \bar{x}_3$ | $(x_2 x_3)$ |
| 11 | $x_3 \vee x_4$ | $(1 - x_3 - x_4 + x_3 x_4)$ |
| 12 | $\bar{x}_3 \vee \bar{x}_4$ | $(x_3 x_4)$ |

Adding the individual clause penalties together gives our QUBO model

$$\min \quad y = 3 + x_1 - 2x_4 - x_2 x_3 + x_2 x_4 + 2x_3 x_4$$

or,

$$\min \quad y = 3 + x^t Q x$$

where the Q matrix is given by

$$\begin{bmatrix} 1 & 0 & 0 & 0 \\ 0 & 0 & -1/2 & 1/2 \\ 0 & -1/2 & 0 & 1 \\ 0 & 1/2 & 1 & -2 \end{bmatrix}$$

Solving QUBO gives: $y = 3 - 2 = 1$ at $x_1 = x_2 = x_3 = 0, x_4 = 1$, meaning that all clauses but one are satisfied.

**Remarks:** The QUBO approach illustrated above has been successfully used in Kochenberger, et. al. (2005) to solve Max 2-sat problems with hundreds of variables and thousands of clauses. An interesting feature of this approach for solving Max 2-sat problems is that the size of the resulting QUBO model to be solved is independent of the number of clauses in the problem and is determined only by the number of variables at hand. Thus, a Max 2-Sat problem with 200 variables and 30,000 clauses can be modeled and solved as a QUBO model with just 200 variables.

## Section 5: Creating QUBO Models: A General Purpose Approach

In this section, we illustrate how to construct an appropriate QUBO model in cases where a QUBO formulation doesn't arise naturally (as we saw in section 3) or where useable penalties are not known in advance (as we saw in section 4). It turns out that for these more general cases, we can always "discover" useable penalties by adopting the procedure outlined below.

For this purpose, consider the <u>general 0/1</u> optimization problem of the form:

$$\min y = x'Cx$$
$$Ax = b, \ x \ binary$$

This model accommodates both quadratic and linear objective functions since the linear case results when C is a diagonal matrix (observing that $x_j^2 = x_j$ when $x_j$ is a 0-1 variable). Under the assumption that A and b have integer components, problems with inequality constraints can always be put in this form by including slack variables and then representing the slack variables by a binary expansion. (For example, this would introduce a slack variable s to convert the inequality $4x_1 + 5x_2 - x_3 \le 6$ into $4x_1 + 5x_2 - x_3 + s = 6$, and since clearly $s \le 7$ (in case $x_3 = 1$), $s$ could be represented by the binary expansion $s_1 + 2s_2 + 4s_3$ where $s_1, s_2$ and $s_3$ are additional binary variables. If it is additionally known that at not both $x_1$ and $x_2$ can be 0, then $s$ can be at most 3 and can be represented by the expansions $s_1 + 2s_2$. A fuller treatment of slack variables is given subsequently.) These constrained quadratic optimization models are converted into equivalent unconstrained QUBO models by converting the constraints $Ax = b$ (representing slack variables as x variables) into quadratic penalties to be added to the objective function, following the same re-casting as we illustrated in section 4.

Specifically, for a positive scalar P, we add a quadratic penalty $P(Ax - b)^t(Ax - b)$ to the objective function to get

$$y = x'Cx + P\left(Ax - b\right)^t \left(Ax - b\right)$$
$$= x'Cx + x'Dx + c$$
$$= x'Qx + c$$

where the matrix D and the additive constant c result directly from the matrix multiplication indicated.  Dropping the additive constant, the equivalent unconstrained version of the constrained problem becomes

$$QUBO : \min x^t Q x, \, x \text{ binary}$$

**Remarks**:

1. A suitable choice of the penalty scalar P, as we commented earlier, can always be chosen so that the optimal solution to QUBO is the optimal solution to the original constrained problem. Solutions obtained can always be checked for feasibility to confirm whether or not appropriate penalty choices have been made.

2. For ease of reference, the preceding procedure that transforms the general problem into an equivalent QUBO model will be called **Transformation # 1**. The mechanics of Transformation #1 can be employed whenever we need to convert linear constraints of the form $Ax = b$ into usable quadratic penalties in our efforts to re-cast a given problem with equality constraints into the QUBO form.  Boros and Hammer (2002) give a discussion of this approach which is the basis for establishing the generality of QUBO. For realistic applications, a program will need to be written implementing Transformation # 1 and producing the Q matrix needed for the QUBO model. Any convenient language, like C++, Python, Matlab, etc., can be used for this purpose.  For small problems, or for preliminary tests preceding large-scale applications, we can usually proceed manually as we'll do in these notes.

3. Note that the additive constant, c, does not impact the optimization and can be ignored during the optimization process. Once the QUBO model has been solved, the constant c can be used to recover the original objective function value.  Alternatively, the original objective function value can always be determined by using the optimal $x_j$ found when QUBO is solved.

Transformation #1 is the "go to" approach in cases where appropriate quadratic penalty functions are not known in advance. In general, it represents an approach that can be adopted for any problem. Due to this generality, Transformation # 1 has proven to be an important modeling tool in many problem settings.

Before moving on to applications in this section, we want to single out another constraint/penalty pair for special recognition that we worked with before in section 4:

$$(x_i + x_j \leq 1) \rightarrow P(x_i x_j)$$

Constraints of this form appear in many important applications. Due to their importance and frequency of use, we refer to this special case as Transformation #2. We'll have occasion to use this as well as Transformation # 1 later in this section.

## 5.1 Set Partitioning

The set partitioning problem (SPP) has to do with partitioning a set of items into subsets so that each item appears in exactly one subset and the cost of the subsets chosen is minimized. This problem appears in many settings including the airline and other industries and is traditionally formulated in binary variables as

$$\min \sum_{j=1}^{n} c_j x_j$$

$$st$$

$$\sum_{j=1}^{n} a_{ij} x_j = 1 \quad for\ i = 1,...m$$

Where $x_j$ denotes whether or not subset j is chosen, $c_j$ is the cost of subset j, and the $a_{ij}$ coefficients are 0 or 1 denoting whether or not variable $x_j$ explicitly appears in constraint i. Note that his model has the form of the general model given at the beginning of this section where, in this case, the objective function matrix C is a diagonal matrix with all off-diagonal elements equal to zero and the diagonal elements are given by the original linear objective function coefficients. Thus, we can re-cast the model into a QUBO model directly by using Transformation # 1. We illustrate this with the following example.

**Numerical Example:** Consider a set partitioning problem

$$\min y = 3x_1 + 2x_2 + x_3 + x_4 + 3x_5 + 2x_6$$

subject to

$$x_1 + x_3 + x_6 = 1$$
$$x_2 + x_3 + x_5 + x_6 = 1$$
$$x_3 + x_4 + x_5 = 1$$
$$x_1 + x_2 + x_4 + x_6 = 1$$

and x binary. Normally, Transformation # 1 would be embodied in a supporting computer routine and employed to re-cast this problem into an equivalent instance of a QUBO model. For this small example, however, we can proceed manually as follows: The conversion to an equivalent QUBO model via Transformation # 1 involves forming quadratic penalties and adding them to the original objective function. In general, the quadratic penalties to be added (for a minimization problem) are given by $P \sum_i (\sum_{j=1}^{n} a_{ij} x_{ij} - b_i)^2$ where the outer summation is taken over all constraints in the system $Ax = b$.

For our example we have

$$\min y = 3x_1 + 2x_2 + x_3 + x_4 + 3x_5 + 2x_6$$
$$+ P(x_1 + x_3 + x_6 - 1)^2 + P(x_2 + x_3 + x_5 + x_6 - 1)^2$$
$$+ P(x_3 + x_4 + x_5 - 1)^2 + P(x_1 + x_2 + x_4 + x_6 - 1)^2$$

Arbitrarily taking P to be 10, and recalling that $x_j^2 = x_j$ since our variables are binary, this becomes

$$\min y = -17x_1{}^2 - 18x_2{}^2 - 29x_3{}^2 - 19x_4{}^2 - 17x_5{}^2 - 28x_6{}^2 + 20x_1x_2 + 20x_1x_3 + 20x_1x_4 + 40x_1x_6$$
$$+ 20x_2x_3 + 20x_2x_4 + 20x_2x_5 + 40x_2x_6 + 20x_3x_4 + 40x_3x_5 + 40x_3x_6 + 20x_4x_5$$
$$+ 20x_4x_6 + 20x_5x_6 + 40$$

Dropping the additive constant 40, we then have our QUBO model

$$\min x^t Q x, \ x \, binary$$

where the Q matrix is

$$Q = \begin{bmatrix} -17 & 10 & 10 & 10 & 0 & 20 \\ 10 & -18 & 10 & 10 & 10 & 20 \\ 10 & 10 & -29 & 10 & 20 & 20 \\ 10 & 10 & 10 & -19 & 10 & 10 \\ 0 & 10 & 20 & 10 & -17 & 10 \\ 20 & 20 & 20 & 10 & 10 & -28 \end{bmatrix}$$

Solving this QUBO formulation gives an optimal solution $x_1 = x_5 = 1$ (with all other variables equal to 0) to yield $y = 6$.

**Remarks**:

1. The QUBO approach to solving set partitioning problems has been successfully applied in Lewis, et. al. (2008) to solve large instances with thousands of variables and hundreds of constraints.

2. The special nature of the set partitioning model allows an alternative to Transformation #1 for constructing the QUBO model. Let $k_j$ denote the number of 1's in the jth column of the constraint matrix A and let $r_{ij}$ denote the number of times variables i and j appear in the same constraint. Then the diagonal elements of Q are given by $q_{ii} = c_i - Pk_i$ and the off – diagonal elements of Q are given by $q_{ij} = q_{ji} = Pr_{ij}$. The additive constant is given by $m * P$. These relationships make it easy to formulate the QUBO model for any set partitioning problem without having to go through the explicit algebra of Transformation # 1.

3. The set partitioning problem may be viewed as a form of clustering problem and is elaborated further in Section 6.

## 5.2 Graph Coloring

In many applications, Transformation # 1 and Transformation # 2 can be used in concert to produce an equivalent QUBO model, as demonstrated next in the context of graph coloring. Vertex coloring problems seek to assign colors to nodes of a graph in such a way that adjacent nodes receive different colors. The K-coloring problem attempts to find such a coloring using exactly K colors. A wide range of applications, ranging from frequency assignment problems to printed circuit board design problems, can be represented by the K-coloring model.

These problems can be modeled as satisfiability problems as follows:

Let $x_{ij} = 1$ if node i is assigned color j, and 0 otherwise.

Since each node must be colored, we have the constraints

$$\sum_{j=1}^{K} x_{ij} = 1 \quad i = 1, ..., n$$

where n is the number of nodes in the graph. A feasible coloring, in which adjacent nodes are assigned different colors, is assured by imposing the constraints

$$x_{ip} + x_{jp} \leq 1 \quad p = 1, ..., K$$

for all adjacent nodes (i,j) in the graph.

This problem, then, can be re-cast in the form of a QUBO model by using Transformation # 1 on the node assignment constraints and using Transformation # 2 on the adjacency constraints. This problem does not have an objective function in its original formulation, meaning our focus is on finding a feasible coloring using the K colors allowed. As a result, any positive value for the penalty P will do. (The resulting QUBO model of course has an objective function given by $x^t Q x$ where Q is determined by the foregoing re-formulation.)

**Numerical Example:** Consider the problem of finding a feasible coloring of the following graph using K= 3 colors.

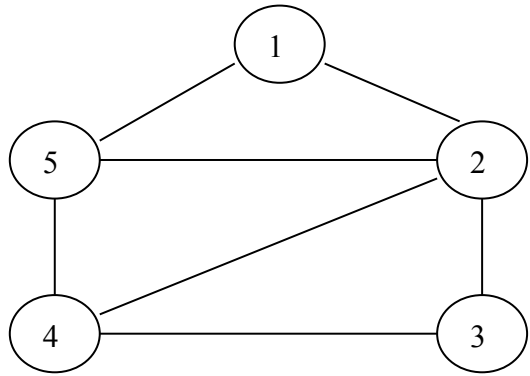

Given the discussion above, we see that the goal is to find a solution to the system:

$$x_{i1} + x_{i2} + x_{i3} = 1 \quad i = 1, 5$$
$$x_{ip} + x_{jp} \leq 1 \quad p = 1, 3$$

(for all adjacent nodes i and j)

In this traditional form, the model has 15 variables and 26 constraints. As suggested above, to recast this problem into the QUBO form, we can use Transformation # 1 on the node assignment equations and Transformation # 2 on adjacency inequalities. One way to proceed here is to start with a 15-by-15 Q matrix where initially all the elements are equal to zero and then re-define appropriate elements based on the penalties obtained from Transformations # 1 and # 2. To clarify the approach, we'll take these two sources of penalties one at a time. For ease of notation and to be consistent with earlier applications, we'll first re-number the variables using a single subscript, from 1 to 15, as follows:

$$\left( x_{11}, x_{12}, x_{13}, x_{21}, x_{22}, x_{23}, x_{31}, \ldots x_{52}, x_{53} \right) = \left( x_1, x_2, x_3, x_4, x_5, x_6, x_7, \ldots x_{14}, x_{15} \right)$$

As we develop our QUBO model, we'll use the variables with a single subscript. First, we'll consider the node assignment equations and the penalties we get from Transformation # 1. Taking these equations in turn we have $P(x_1 + x_2 + x_3 - 1)^2$ which becomes $P(-x_1 - x_2 - x_3 + 2x_1x_2 + 2x_1x_3 + 2x_2x_3) + P$.

$P(x_4 + x_5 + x_6 - 1)^2$ which becomes $P(-x_4 - x_5 - x_6 + 2x_4x_5 + 2x_4x_6 + 2x_5x_6) + P$.

$P(x_7 + x_8 + x_9 - 1)^2$ which becomes $P(-x_7 - x_8 - x_9 + 2x_7x_8 + 2x_7x_9 + 2x_8x_9) + P$.

$P(x_{10} + x_{11} + x_{12} - 1)^2$ which becomes $P(-x_{10} - x_{11} - x_{12} + 2x_{10}x_{11} + 2x_{10}x_{12} + 2x_{11}x_{12}) + P$.

$P(x_{13} + x_{14} + x_{15} - 1)^2$ which becomes $P(-x_{13} - x_{14} - x_{15} + 2x_{13}x_{14} + 2x_{13}x_{15} + 2x_{14}x_{15}) + P$.

Taking P to equal 4 and inserting these penalties in the "developing" Q matrix gives the following partially completed Q matrix along with an additive constant of 5P.

$$
\begin{bmatrix}
-4 & 4 & 4 & 0 & 0 & 0 & 0 & 0 & 0 & 0 & 0 & 0 & 0 & 0 & 0 \\
4 & -4 & 4 & 0 & 0 & 0 & 0 & 0 & 0 & 0 & 0 & 0 & 0 & 0 & 0 \\
4 & 4 & -4 & 0 & 0 & 0 & 0 & 0 & 0 & 0 & 0 & 0 & 0 & 0 & 0 \\
0 & 0 & 0 & -4 & 4 & 4 & 0 & 0 & 0 & 0 & 0 & 0 & 0 & 0 & 0 \\
0 & 0 & 0 & 4 & -4 & 4 & 0 & 0 & 0 & 0 & 0 & 0 & 0 & 0 & 0 \\
0 & 0 & 0 & 4 & 4 & -4 & 0 & 0 & 0 & 0 & 0 & 0 & 0 & 0 & 0 \\
0 & 0 & 0 & 0 & 0 & 0 & -4 & 4 & 4 & 0 & 0 & 0 & 0 & 0 & 0 \\
0 & 0 & 0 & 0 & 0 & 0 & 4 & -4 & 4 & 0 & 0 & 0 & 0 & 0 & 0 \\
0 & 0 & 0 & 0 & 0 & 0 & 4 & 4 & -4 & 0 & 0 & 0 & 0 & 0 & 0 \\
0 & 0 & 0 & 0 & 0 & 0 & 0 & 0 & 0 & -4 & 4 & 4 & 0 & 0 & 0 \\
0 & 0 & 0 & 0 & 0 & 0 & 0 & 0 & 0 & 4 & -4 & 4 & 0 & 0 & 0 \\
0 & 0 & 0 & 0 & 0 & 0 & 0 & 0 & 0 & 4 & 4 & -4 & 0 & 0 & 0 \\
0 & 0 & 0 & 0 & 0 & 0 & 0 & 0 & 0 & 0 & 0 & 0 & -4 & 4 & 4 \\
0 & 0 & 0 & 0 & 0 & 0 & 0 & 0 & 0 & 0 & 0 & 0 & 4 & -4 & 4 \\
0 & 0 & 0 & 0 & 0 & 0 & 0 & 0 & 0 & 0 & 0 & 0 & 4 & 4 & -4 \\
\end{bmatrix}
$$

Note the block diagonal structure. Many problems have patterns that can be exploited in developing Q matrices needed for their QUBO representation. Looking for patterns is often a useful de-bugging tool.

To complete our Q matrix, it's a simple matter of inserting the penalties representing the adjacency constraints into the above matrix. For these, we use the penalties of Transformation # 2, namely $Px_ix_j$, for each adjacent pair of nodes and each of the three allowed colors. We have 7 adjacent pairs of nodes and three colors, yielding a total of 21 adjacency constraints. Allowing for symmetry, we'll insert 42 penalties into the matrix, augmenting the penalties already in place. For example, for the constraint ensuring that nodes 1 and 2 can not both have color #1, the penalty is $Px_1x_4$, implying that we insert the penalty value "2" in row 1 and column 4 of our matrix and also in column 1 and row 4. (Recall that we have relabeled our variables so that the original variables $x_{1,1}$ and $x_{2,1}$ are now variables $x_1$ and $x_4$.) Including the penalties for the other adjacency constraints completes the Q matrix as shown below

$$Q = \begin{bmatrix}
-4 & 4 & 4 & 2 & 0 & 0 & 0 & 0 & 0 & 0 & 0 & 0 & 2 & 0 & 0 \\
4 & -4 & 4 & 0 & 2 & 0 & 0 & 0 & 0 & 0 & 0 & 0 & 0 & 2 & 0 \\
4 & 4 & -4 & 0 & 0 & 2 & 0 & 0 & 0 & 0 & 0 & 0 & 0 & 0 & 2 \\
2 & 0 & 0 & -4 & 4 & 4 & 2 & 0 & 0 & 2 & 0 & 0 & 2 & 0 & 0 \\
0 & 2 & 0 & 4 & -4 & 4 & 0 & 2 & 0 & 0 & 2 & 0 & 0 & 2 & 0 \\
0 & 0 & 2 & 4 & 4 & -4 & 0 & 0 & 2 & 0 & 0 & 2 & 0 & 0 & 2 \\
0 & 0 & 0 & 2 & 0 & 0 & -4 & 4 & 4 & 2 & 0 & 0 & 0 & 0 & 0 \\
0 & 0 & 0 & 0 & 2 & 0 & 4 & -4 & 4 & 0 & 2 & 0 & 0 & 0 & 0 \\
0 & 0 & 0 & 0 & 0 & 2 & 4 & 4 & -4 & 0 & 0 & 2 & 0 & 0 & 0 \\
0 & 0 & 0 & 2 & 0 & 0 & 2 & 0 & 0 & -4 & 4 & 4 & 2 & 0 & 0 \\
0 & 0 & 0 & 0 & 2 & 0 & 0 & 2 & 0 & 4 & -4 & 4 & 0 & 2 & 0 \\
0 & 0 & 0 & 0 & 0 & 2 & 0 & 0 & 2 & 4 & 4 & -4 & 0 & 0 & 2 \\
2 & 0 & 0 & 2 & 0 & 0 & 0 & 0 & 0 & 2 & 0 & 0 & -4 & 4 & 4 \\
0 & 2 & 0 & 0 & 2 & 0 & 0 & 0 & 0 & 0 & 2 & 0 & 4 & -4 & 4 \\
0 & 0 & 2 & 0 & 0 & 2 & 0 & 0 & 0 & 0 & 0 & 2 & 4 & 4 & -4
\end{bmatrix}$$

The above matrix incorporates all of the constraints of our coloring problem, yielding the equivalent QUBO model

$$QUBO : \min x^t Q x$$

Solving this model yields the feasible coloring:
$x_2 = x_4 = x_9 = x_{11} = x_{15} = 1$ with all other variables equal to zero.

Switching back to our original variables, this solution means that nodes 1 and 4 get color #2, node 2 gets color # 1, and nodes 3 and 5 get color # 3.

**Remark**: This approach to graph coloring problems has proven to be very effective for a wide variety of coloring instances with hundreds of nodes, as demonstrated in Kochenberger, et. al. (2005).

## 5.3 General 0/1 Programming

Many important problems in industry and government can be modeled as 0/1 linear programs with a mixture of constraint types. The general problem of this nature can be represented in matrix form by

$$\max cx$$
$$st$$
$$Ax = b$$
$$x \, binary$$

where <u>slack variables</u> are introduced as needed to convert inequality constraints into equalities. Given a problem in this form, Transformation # 1 can be used to re-cast the problem into the QUBO form

$$\max x_0 = x^t Q x$$
$$st \; x \, binary$$

As discussed earlier, problems with inequality constraints can be handled by introducing slack variables, via a binary expansion, to create the system of constraints $Ax = b$.

**Numerical Example**: Consider the general 0/1 problem

$$\max \; 6x_1 + 4x_2 + 8x_3 + 5x_4 + 5x_5$$
$$st$$
$$2x_1 + 2x_2 + 4x_3 + 3x_4 + 2x_5 \leq 7$$
$$1x_1 + 2x_2 + 2x_3 + 1x_4 + 2x_5 = 4$$
$$3x_1 + 3x_2 + 2x_3 + 4x_4 + 4x_5 \geq 5$$
$$x \in \{0,1\}$$

Since Transformation # 1 requires all constraints to be equations rather than inequalities, we convert the 1st and 3rd constraints to equations by including slack variables via a binary expansion. To do this, we first estimate upper bounds on the slack activities as a basis for determining how many binary variables will be required to represent the slack variables in the binary expansions. Typically, the upper bounds are determined simply by examining the constraints and estimating a reasonable value for how large the slack activity could be. For the problem at hand, we can refer to the slack variables for constraints 1 and 3 as $s_1$ and $s_3$ with upper bounds 3 and 6 respectively. Our binary expansions are:

$$0 \leq s_1 \leq 3 \;\; \Rightarrow s_1 = 1x_6 + 2x_7$$
$$0 \leq s_3 \leq 6 \;\; \Rightarrow s_3 = 1x_8 + 2x_9 + 4x_{10}$$

Where $x_6, x_7, x_8, x_9$ and $x_{10}$ are new binary variables. Note that these new variables will have objective function coefficients equal to zero. Including these slack variables gives the system

$Ax = b$ with $A$ given by:

$$A = \begin{bmatrix} 2 & 2 & 4 & 3 & 2 & 1 & 2 & 0 & 0 & 0 \\ 1 & 2 & 2 & 1 & 2 & 0 & 0 & 0 & 0 & 0 \\ 3 & 3 & 2 & 4 & 4 & 0 & 0 & -1 & -2 & -4 \end{bmatrix}$$

We can now use Transformation # 1 to reformulate our problem as a QUBO instance. Adding the penalties to the objective function gives

$$\begin{aligned} \max y = {} & 6x_1 + 4x_2 + 8x_3 + 5x_4 + 5x_5 \\ & - P(2x_1 + 2x_2 + 4x_3 + 3x_4 + 2x_5 + 1x_6 + 2x_7 - 7)^2 \\ & - P(1x_1 + 2x_2 + 2x_3 + 1x_4 + 2x_5 - 4)^2 \\ & - P(3x_1 + 3x_2 + 2x_3 + 4x_4 + 4x_5 - 1x_8 - 2x_9 - 4x_{10} - 5)^2 \end{aligned}$$

Taking P = 10 and re-writing this in the QUBO format gives

$$\max \quad y = x'Qx$$

with an additive constant of -900 and a Q matrix

$$Q = \begin{bmatrix} 526 & -150 & -160 & -190 & -180 & -20 & -40 & 30 & 60 & 120 \\ -150 & 574 & -180 & -200 & -200 & -20 & -40 & 30 & 60 & 120 \\ -160 & -180 & 688 & -220 & -200 & -40 & -80 & 20 & 40 & 80 \\ -190 & -200 & -220 & 645 & -240 & -30 & -60 & 40 & 80 & 160 \\ -180 & -200 & -200 & -240 & 605 & -20 & -40 & 40 & 80 & 160 \\ -20 & -20 & -40 & -30 & -20 & 130 & -20 & 0 & 0 & 0 \\ -40 & -40 & -80 & -60 & -40 & -20 & 240 & 0 & 0 & 0 \\ 30 & 30 & 20 & 40 & 40 & 0 & 0 & -110 & -20 & -40 \\ 60 & 60 & 40 & 80 & 80 & 0 & 0 & -20 & -240 & -80 \\ 120 & 120 & 80 & 160 & 160 & 0 & 0 & -40 & -80 & -560 \end{bmatrix}$$

Solving $\max y = x^t Q x$ gives the non-zero values

$$x_1 = x_4 = x_5 = x_9 = x_{10} = 1$$

for which $y = 916$. Note that the third constraint is loose. Adjusting for the additive constant, it gives an objective function value of 16. Alternatively, we could have simply evaluated the original objective function at the solution $x_1 = x_4 = x_5 = 1$ to get the objective function value of 16.

**Remarks**: Any problem in linear constraints and bounded integer variables can be converted through a binary expansion into max $y = x^t Q x$ as illustrated here. In such applications, however, the elements of the Q matrix can, depending on the data, get unacceptably large and may require suitable scaling to mitigate this problem.

## 5.4 Quadratic Assignment

The Quadratic Assignment Problem (QAP) is a renowned problem in combinatorial optimization with applications in a wide variety of settings. It is also one of the more challenging models to solve. The problem setting is as follows: We are given n facilities and n locations along with a flow matrix ($f_{ij}$) denoting the flow of material between facilities i and j. A distance matrix ($d_{ij}$) specifies the distance between sites i and j. The optimization problem is to find an assignment of facilities to locations to minimize the weighted flow across the system. Cost information can be explicitly introduced to yield a cost minimization model, as is common in some applications. The decision variables are $x_{ij} = 1$ if facility i is assigned to location j; otherwise, $x_{ij} = 0$. Then the classic QAP model can be stated as:

$$\text{Minimize} \sum_{i=1}^{n} \sum_{j=1}^{n} \sum_{k=1}^{n} \sum_{l=1}^{n} f_{ij} d_{kl} x_{ik} x_{jl}$$

Subject to
$$\sum_{i=1}^{n} x_{ij} = 1 \quad j = 1, n$$

$$\sum_{j=1}^{n} x_{ij} = 1 \quad i = 1, n$$

$$x_{ij} \in \{0,1\}, \quad i, j = 1, n$$

All QAP problems have $n^2$ variables, which often yields large models in practical settings. This model has the general form presented at the beginning of this section and consequently Transformation # 1 can be used to convert any QAP problem into a QUBO instance.

**Numerical Example**: Consider a small example with n = 3 facilities and 3 locations with flow and distance matrices respectively given as follows:

$$\begin{bmatrix} 0 & 5 & 2 \\ 5 & 0 & 3 \\ 2 & 3 & 0 \end{bmatrix} \text{ and } \begin{bmatrix} 0 & 8 & 15 \\ 8 & 0 & 13 \\ 15 & 13 & 0 \end{bmatrix}.$$

It is convenient to re-label the variables using only a single subscript as we did previously in the graph coloring problem, thus replacing

$$(x_{11}, x_{12}, x_{13}, x_{21}, x_{22}, x_{23}, x_{31}, x_{32}, x_{33}) \text{ by } (x_1, x_2, x_3, x_4, x_5, x_6, x_7, x_8, x_9)$$

Given the flow and distance matrices our QAP model becomes:

$$\min x_0 = 80x_1x_5 + 150x_1x_6 + 32x_1x_8 + 60x_1x_9 + 80x_2x_4 + 130x_2x_6 + 60x_2x_7 + 52x_2x_9$$
$$+ 150x_3x_4 + 130x_3x_5 + 60x_3x_7 + 52x_3x_8 + 48x_4x_8 + 90x_4x_9 + 78x_5x_9 + 78x_6x_8$$

subject to
$$x_1 + x_2 + x_3 = 1$$
$$x_4 + x_5 + x_6 = 1$$
$$x_7 + x_8 + x_9 = 1$$
$$x_1 + x_4 + x_7 = 1$$
$$x_2 + x_5 + x_8 = 1$$
$$x_3 + x_6 + x_9 = 1$$

Converting the constraints into quadratic penalty terms and adding them to the objective function gives the unconstrained quadratic model

$$\min y = 80x_1x_5 + 150x_1x_6 + 32x_1x_8 + 60x_1x_9 + 80x_2x_4 + 130x_2x_6 + 60x_2x_7 + 52x_2x_9$$
$$+ 150x_3x_4 + 130x_3x_5 + 60x_3x_7 + 52x_3x_8 + 48x_4x_8 + 90x_4x_9 + 78x_5x_9 + 78x_6x_8$$
$$+ P(x_1 + x_2 + x_3 - 1)^2 + P(x_4 + x_5 + x_6 - 1)^2 + P(x_7 + x_8 + x_9 - 1)^2$$
$$+ P(x_1 + x_4 + x_7 - 1)^2 + P(x_2 + x_5 + x_8 - 1)^2 + P(x_3 + x_6 + x_9 - 1)^2$$

Choosing a penalty value of P = 200, this becomes the standard QUBO problem

$$\text{QUBO:} \quad \min \quad y = x^t Q x$$

with an additive constant of 1200 and the following 9-by-9 Q matrix:

$$\begin{bmatrix}
-400 & 200 & 200 & 200 & 40 & 75 & 200 & 16 & 30 \\
200 & -400 & 200 & 40 & 200 & 65 & 16 & 200 & 26 \\
200 & 200 & -400 & 75 & 65 & 200 & 30 & 26 & 200 \\
200 & 40 & 75 & -400 & 200 & 200 & 200 & 24 & 45 \\
40 & 200 & 65 & 200 & -400 & 200 & 24 & 200 & 39 \\
75 & 65 & 200 & 200 & 200 & -400 & 45 & 39 & 200 \\
200 & 16 & 30 & 200 & 24 & 45 & -400 & 200 & 200 \\
16 & 200 & 26 & 24 & 200 & 39 & 200 & -400 & 200 \\
30 & 26 & 200 & 45 & 39 & 200 & 200 & 200 & -400
\end{bmatrix}$$

Solving QUBO gives $y = -982$ at $x_1 = x_5 = x_9 = 1$ and all other variables = 0. Adjusting for the additive constant, we get the original objective function value of 1200 -982 =218.

**Remark**: A QUBO approach to solving QAP problems, as illustrated above, has been successfully applied to problems with more than 30 facilities and locations in Wang, et. al. (2016).

## 5.5 Quadratic Knapsack

Knapsack problems, like the other problems presented earlier in this section, play a prominent role in the field of combinatorial optimization, having widespread application in such areas as project selection and capital budgeting. In such settings, a set of attractive potential projects is identified and the goal is to identify a subset of maximum value (or profit) that satisfies the budget limitations. The classic linear knapsack problem applies when the value of a project depends only on the individual projects under consideration. The quadratic version of this problem arises when there is an interaction between pairs of projects affecting the value obtained.

For the general case with n projects, the Quadratic Knapsack Problem (QKP) is commonly modeled as

$$\max \sum_{i=1}^{n-1} \sum_{j=i}^{n} v_{ij} x_i x_j$$

subject to the budget constraint

$$\sum_{j=1}^{n} a_j x_j \leq b$$

Where $x_j = 1$ if project j is chosen: else, $x_j = 0$. The parameters $v_{ij}, a_j$ and $b$ represent, respectively, the value associated with choosing projects i and j, the resource requirement of project j, and the total resource budget. Generalizations involving multiple knapsack constraints are found in a variety of application settings.

**Numerical Example**: Consider the QKP model with four projects:

$$\max 2x_1 + 5x_2 + 2x_3 + 4x_4 + 8x_1x_2 + 6x_1x_3 +$$
$$10x_1x_4 + 2x_2x_3 + 6x_2x_4 + 4x_3x_4$$

subject to the knapsack constraint:

$$8x_1 + 6x_2 + 5x_3 + 3x_4 \leq 16$$

We re-cast this into the form of a QUBO model by first converting the constraint into an equation and then using the ideas embedded in Transformation # 1. Introducing a slack variable in the form of the binary expansion $1x_5 + 2x_6$, we get the equality constraint

$$8x_1 + 6x_2 + 5x_3 + 3x_4 + 1x_5 + 2x_6 = 16$$

which we can convert to penalties to produce our QUBO model as follows.

Including the penalty term in the objective function gives the unconstrained quadratic model:

$$\max y = 2x_1 + 5x_2 + 2x_3 + 4x_4 + 8x_1x_2 + 6x_1x_3$$
$$+ 10x_1x_4 + 2x_2x_3 + 6x_2x_4 + 4x_3x_4$$
$$- P(8x_1 + 6x_2 + 5x_3 + 3x_4 + 1x_5 + 2x_6 - 16)^2$$

Choosing a penalty P = 10, and cleaning up the algebra gives the QUBO model

$$\text{QUBO: } \max y = x^t Q x$$

with an additive constant of -2560 and the Q matrix

$$\begin{bmatrix} 1922 & -476 & -397 & -235 & -80 & -160 \\ -476 & 1565 & -299 & -177 & -60 & -120 \\ -397 & -299 & 1352 & -148 & -50 & -100 \\ -235 & -177 & -148 & 874 & -30 & -60 \\ -80 & -60 & -50 & -30 & 310 & -20 \\ -160 & -120 & -100 & -60 & -20 & 600 \end{bmatrix}$$

Solving QUBO gives $y = 2588$ at $x = (1,0,1,1,0,0)$. Adjusting for the additive constant, gives the value 28 for the original objective function.

**Remark**: The QUBO approach to QKP has proven to be successful on problems with several hundred variables as shown in Glover, et. al. (2002).

## Section 6: Connections to Quantum Computing and Machine Learning

*Quantum Computing QUBO Developments:* -- As noted in Section 1, one of the most significant applications of QUBO emerges from its equivalence to the famous Ising problem in physics. In common with the earlier demonstration that a remarkable array of NP-hard problems can converted into the QUBO form, Lucas (2014) more recently has observed that such problems can be converted into the Ising form, including graph and number partitioning, covering and set packing, satisfiability, matching, and constrained spanning tree problems, among others. Pakin (2017) presents an algorithm for finding the shortest path through a maze by expressing the shortest path as the globally optimal value of an Ising Hamiltonian instead of via a traditional backtracking mechanism. Ising problems replace $x \in \{0, 1\}$n by $x \in \{-1, 1\}$ n and can be put in the QUBO form by defining xj' = (xj + 1)/2 and then redefining xj to be xj'.[1] Efforts to solve Ising problems are often carried out with annealing approaches, motivated by the perspective in physics of applying annealing methods to find a lowest energy state.

More effective methods for QUBO problems, and hence for Ising problems, are obtained using modern metaheuristics. Among the best metaheuristic methods for QUBO are those based on tabu search and path relinking as described in Glover (1996, 1997), Glover and Laguna (1997) and adapted to QUBO in Wang et al. (2012, 2013).
A bonus from this development has been to create a link between QUBO problems and quantum computing.[2] A quantum computer based on quantum annealing with an integrated physical network structure of qubits known as a Chimera graph has incorporated ideas from Wang et al. (2012) in its software and has been implemented on the D-Wave System. The ability to obtain a quantum speedup effect for this system applied to QUBO problems has been demonstrated in Boixo et al. (2014).

Additional advances incorporating methodology from Wang et al. (2012, 2013) are provided in the D-Wave open source software system Qbsolv (2017) and in the supplementary QMASM system by Pakin (2018). Qbsolv is a hyrid classical/hardware accelerator tool, which takes as input a QUBO that may be larger/denser/higher-precision than the accelerator, and solves subQUBOs on an accelerator and combines the results for full QUBO solution. It has enabled widespread experimentation to map optimization problems to the QUBO form for execution on classical and D-wave computers. D-Wave has now upgraded this system by drawing on the MIT Kerberos system (Kerberos, 2019) which offers many convenience features for users. The Quantum Bridge Analytics perspective, as elaborated below, is providing additional gains.

---

[1] This adds a constant to (1), which is irrelevant for optimization.

[2] Reference to quantum computing would not be complete without mentioning Google's recent claim to achieving 'quantum supremacy.' This outcome has no bearing on the computational considerations discussed here. See, for example, Preskill (2019).

Recent QUBO quantum computing applications, complementing earlier applications on classical computing systems, include those for graph partitioning problems in Mniszewski et al. (2016) and Ushijima-Mwesigwa et al. (2017); graph clustering (quantum community detection problems) in Negre et al. (2018, 2019); traffic-flow optimization in Neukart et al. (2017); vehicle routing problems in Feld et al. (2018), Clark et al. (2019) and Ohzeki et al.(2018); maximum clique problems in Chapuis et al. (2018); cybersecurity problems in Munch et al. (2018) and Reinhardt et al.(2018); predictive health analytics problems in De Oliveira et al. (2018) and Sahner et al. (2018); and financial portfolio management problems in Elsokkary et al. (2017) and Kalra et al. (2018). In another recent development, QUBO models are being studied using the IBM neuromorphic computer at as reported in Alom et al. (2017) and Aimone et al. (2018). Still more recently, Aramon, et al. (2019) have investigated and tested the Fujitsu Digital Annealer approach, which is also designed to solve fully connected QUBO problems, implemented on application-specific CMOS hardware and solved problems of 1,024 variables.

Multiple quantum computational paradigms are emerging as important research topics, and their relative merits have been the source of some controversy. One of the most active debates concerns the promise of quantum gate systems, also known as quantum circuit systems, versus the promise of adiabatic or quantum annealing systems.  Part of this debate has concerned the question of whether adiabatic quantum computing incorporates the critical element of quantum entanglement. After some period, the debate was finally resolved by Albash et al. (2015) and Lanting et al. (2015), demonstrating that this question can be answered in the affirmative. Yet another key consideration involves the role of decoherence. Some of the main issues are discussed in Amin et al. (2008) and Albash and Lidar (2015). The challenge is for the gate model to handle decoherence effectively. Superconducting qubit techniques have very short-lived coherence times and the adiabatic approach does not require them, while the gate model does. An important discovery by Yu et al. (2018) shows that the adiabatic and gate systems offer effectively the same potential for achieving the gains inherent in quantum computing processes, with a mathematical demonstration that the quantum circuit algorithm can be transformed into the quantum adiabatic algorithm with the exact same time complexity. This has useful implications for the relevance of QUBO models that have been implemented in an adiabatic quantum annealing setting, disclosing that analogous advances associated with QUBO models may ultimately be realized through quantum circuit systems.

Complementing this analysis, Shaydulin et al. (2018) have conducted a first performance comparison of these two leading paradigms, showing that quantum local search approach with both frameworks can achieve results comparable to state-of-the-art local search using classical computing architectures, with a potential for the quantum approaches to outperform the classical systems as hardware evolves. However, the time frame for realizing such potential has been estimated by some analysts to lie 10 or more years in the future (Reedy, 2017; Debenedictis, 2019).

Regardless of which quantum paradigm proves superior (and when this paradigm will become competitive with the best classical computing systems), the studies of Alom et al. (2017) and Aimone et al. (2018) in neuromorphic computing reinforce the studies of adiabatic and gate based models by indicating the growing significance of the QUBO/Ising model across multiple frameworks.

However, to set the stage for solving QUBO problems on quantum computers, these problems must be embedded (or compiled) onto quantum computing hardware, which in itself is a very hard problem. Date et al. (2019) address this issue by proposing an efficient algorithm for embedding QUBO problems that runs fast, uses less qubits than previous approaches and gets an objective function value close to the global minimum value. In a computational comparison, they find that their embedding algorithm outperforms the embedding algorithm of D-Wave, which is the current state of the art.

Vyskocil et al. (2019) observe that the transformation in Section 5.3 for handling general inequality constraints of the form $\sum_{i=1}^{n} x_i \leq k$ introduces penalties for numerous cross products, which poses difficulties for current quantum annealers such as those by D-Wave Systems. The authors give a scalable and modular two-level approach for handling this situation that first solves a small preliminary mixed integer optimization problem with 16 binary variables and 16 constraints, and then uses this to create a transformation that increases the number of QUBO variables but keeps the number of cross product terms in check, thereby aiding a quantum computer implementation.

Nevertheless, other considerations are relevant for evaluating the performance of different computational paradigms for solving QUBO problems, among them the use of reduction and preprocessing methods for decomposing large scale QUBO problem instances into smaller ones. Hahn et al. (2017) and Pelofske et al. (2019) investigate such preprocessing methods that utilize upper and lower bound heuristics in conjunction with graph decomposition, vertex and edge extraction and persistency analysis. Additional preprocessing methods are introduced in Glover et al. (2018) as described subsequently in the context of machine learning.

*Quantum Bridge Analytics: Joining Classical and Quantum Computing Paradigms:*-- As emphasized in the 2019 Consensus Study Report titled Quantum Computing: Progress and Prospects, by the National Academies of Sciences, Engineering and Medicine (2019), quantum computing will remain in its infancy for some years to come, and in the interim "formulating an R&D program with the aim of developing commercial applications for near-term quantum computing is critical to the health of the field." As noted in this report, such a program will rest on developing "hybrid classical-quantum techniques," which is the focus of Quantum Bridge Analytics. With the emergence of Quantum Bridge Analytics (QBA), a field devoted to bridging the gap between classical and quantum computational methods and technologies, the creation of effective foundations for such hybrid systems is being actively pursued with the development of the Alpha-QUBO solver (2019). This work is paving the way for a wide range of additional QUBO and QUBO-related applications in commercial and academic research settings. The power of the QBA approach has recently been demonstrated in Glover and Kochenberger (2019), with computational tests showing that a relative of Alpha-QUBO, called QUBO 2.0, solves QUBO problems between 100 and 500 variables up to three orders of magnitude faster than a mainstream quantum computing system using Kerberos, and is additionally capable of solving much larger problems involving many thousands of variables.

Another blend of classical and quantum computing, known as the Quantum Approximate Optimization Algorithm (QAOA), is a hybrid variational algorithm introduced by Farhi et al. (2014) that produces approximate solutions for combinatorial optimization problems. The QAOA approach has been recently been applied in Zhou et al. (2018) to MaxCut (MC) problems, including a variant in process for Max Independent Set (MIS) problems, and is claimed by its authors to have the potential to challenge the leading classical algorithms. In theory, QAOA methods can be applied to more types of combinatorial optimization problems than embraced by the QUBO model, but at present the MC and MIS problems studied by QAOA are a very small segment of the QUBO family and no time frame is offered for gaining the ability to tackle additional QUBO problem instances. Significantly, the parameters of the QAOA framework must be modified to produce different algorithms to appropriately handle different problem types. Whether this may limit the universality of this approach in a practical sense remains to be seen.

Wang and Abdullah (2018) acknowledge that the acclaim given to QAOA for exhibiting the feature called "quantum supremacy" does not imply that QAOA will be able to outperform classical algorithms on important combinatorial optimization problems such as Constraint Satisfaction Problems, and current implementations of QAOA are subject to a gate fidelity limitation, where the potential advantages of larger values of the parameter p in QAOA applications are likely to be countered by a decrease in solution accuracy.

QAOA has inspired many researchers to laud its potential virtues, though the practical significance of this potential at present is not well established.   Investigations are currently underway in Kochenberger et al. (2019) to examine this issue by computational testing on a range of QUBO models that fall within the scope of QAOA implementations presently available, to determine the promise of QAOA in relation to classical optimization on these models.
We now examine realms of QUBO models that are actively being investigated apart from issues of alternative computational frameworks for solving them efficiently.

*Unsupervised Machine Learning with QUBO*: -- One of the most salient forms of unsupervised machine learning is represented by clustering. The QUBO set partitioning model provides a very natural form of clustering and gives this model a useful link to unsupervised machine learning. As observed in Ailon et al.(2008) and Aloise, et al.(2013), the CPP (clique partitioning problem) is popular in the area of machine learning as it offers a general model for correlation clustering (CC) and the modularity maximization (MM). Pudenz and Lidar (2013) further show how a QUBO based quantum computing model can be used in unsupervised machine learning. A related application in O'Malley et al. (2018) investigates nonnegative/binary matrix factorization with a D-Wave quantum annealer.

An application of QUBO to unsupervised machine learning in Glover et al. (2018) provides an approach that can be employed either together with quantum computing or independently. In a complementary development, clustering is used to facilitate the solution of QUBO models in Samorani et al. (2018), thereby providing a foundation for studying additional uses of clustering in this context.

*Supervised Machine Learning with QUBO:* -- A proposal to use QUBO in supervised machine learning is introduced in Schneidman, et al. (2006). From the physics perspective, the authors argue that the equivalent Ising model is useful for any representation of neural function, based on the supposition that a statistical model for neural activity should be chosen using the principle of maximum entropy. Consequently, this model has a natural role in statistical neural models of supervised machine learning. Hamilton et al. (2018) discussed the potential to use advance computing such as neuromorphic processing units and quantum annealers in spin-glass networks, Boltzmann machines, convolutional neural networks and constraint satisfaction problems.

*Machine Learning to Improve QUBO Solution Processes:* -- The development of rules and strategies to learn the implications of specific model instances has had a long history. Today this type of machine learning permeates the field of mixed integer programming to identify relationships such as values (or bounds) that can be assigned to variables, or inequalities that can constrain feasible spaces more tightly. Although not traditionally viewed through the lens of machine learning, due in part to being classified under the name of pre-processing, these approaches are now widely acknowledged to constitute a viable and important part of the machine learning domain.

Efforts to apply machine learning to uncover the implications of QUBO problem structures have proceeded more slowly than those devoted to identifying such implications in the mixed integer programming field. A landmark paper in the QUBO area is the work of Boros et al. (2008), which uses roof duality and a max-flow algorithm to provide useful model inferences. More recently, sets of logical tests have been developed in Glover et al. (2018) to learn relationships among variables in QUBO applications which achieved a 45% reduction in size for about half of the problems tested, and in 10 cases succeeded in fixing all the variables, exactly solving these problems. The rules also identified implied relationships between pairs of variables that resulted in simple logical inequalities to facilitate solving these problems.

Other types of machine learning approaches also merit a closer look in the future for applications with QUBO. Among these are the Programming by Optimization approach of Hoos (2012) and the Integrative Population Analysis approach of Glover et al. (1998).

**Section 7: Concluding Remarks**

The benefits of re-casting problems into the QUBO framework, to enable a given binary optimization problem to be solved by a specialized QUBO solver, strongly commend this approach in the remarkable variety of settings where it can be implemented successfully, as illustrated in this tutorial. We conclude by highlighting key ideas relevant to QUBO modeling and its applications in both classical and quantum computing.

1. As previously noted, the National Academies of Sciences, Engineering and Medicine have released a consensus study report on progress and prospects in quantum computing (2019) that discloses the relevance of marrying quantum and classical computing, stating that "formulating an R&D program with the aim of developing commercial applications for near-term quantum computing is critical to the health of the field. Such a program would include … identification of algorithms for which hybrid classical-quantum techniques using modest-size quantum subsystems can provide significant speedup." Studies devoted to this challenge are currently underway at the Los Alamos National Laboratory to investigate the possibilities for achieving such speedup by integrating quantum computing initiatives in conjunction with classical computing approaches such as those embedded in the Alpha-QUBO system (2019).

2. Logical analysis to identify relationships between variables in the work of Glover et al. (2017) can be implemented in the setting of quantum computing to combat the difficulties of applying current quantum computing methods to scale effectively for solving large problems. Approximation methods based on such analysis can be used for decomposing and partitioning large QUBO problems to solve large problems and provide strategies relevant to a broad range of quantum computing applications.

3. In both classical and quantum settings, the transformation to QUBO can sometimes be aided considerably by first employing a change of variables. This is particularly useful in settings where the original model is an edge-based graph model, as in clique partitioning where the standard models can have millions of variables due to the number of edges in the graph. A useful alternative is to introduce node-based variables, by replacing each edge variable with the product of two node variables. Such a change converts a linear model into a quadratic model with many fewer variables, since a graph normally has a much smaller number of nodes than edges. The resulting quadratic model, then, can be converted to a QUBO model by the methods illustrated earlier.

4. Problems involving higher order polynomials arise in certain applications and can be re-cast into a QUBO framework by employing a reduction technique following the ideas of Rosenberg (1975), Rodriques-Heck (2018) and Verma et al.(2019). For example, consider a problem with a cubic term $x_1 x_2 x_3$ in binary variables. Replace the product $x_1 x_2$ by a binary variable, $y_1$ and add a penalty to the objective function of the form $P(x_1 x_2 - 2x_1 y_1 - 2x_2 y_1 + 3y_1)$. By this process, when the optimization drives the penalty term to 0, which happens only when $y_1 = x_1 x_2$, we have reduced the cubic term to an equivalent quadratic term $y_1 y_3$. This procedure can be used recursively to convert higher order polynomials to quadratic models of the QUBO form.

5. The general procedure of Transformation # 1 has similarities to the Lagrange Multiplier approach of classical optimization. The key difference is that our scalar penalties (P) are not "dual" variables to be determined by the optimization. Rather, they are parameters set a priori to encourage the search process to avoid candidate solutions that are infeasible. Moreover, the Lagrange Multiplier approach is not assured to yield a solution that satisfies the problem constraints except in the special case of convex optimization, in contrast to the situation with the QUBO model. To determine good values for Lagrange multipliers (which in general only yield a lower bound instead of an optimum value for the problem objective) recourse must be made to an additional type of optimization called subgradient optimization, which QUBO models do not depend on.

6. Solving QUBO models: Continuing progress in the design and implementation of methods for solving QUBO models will have an impact across a wide range of practical applications of optimization and machine learning. The bibliography that follows gives references to some of the more prominent methods for solving these models.

## Bibliography:

N. Ailon, M. Charikar, A. Newman (2008) "Aggregating inconsistent information: ranking and clustering", *Journal of the ACM (JACM)*, 55(5), 23.

J.B. Aimone, K.E.Hamilton, S. Mniszewsk, L. Reeder, C.D. Schuman, W.M.Severa (2018) "Non-Neural Network Applications for Spiking Neuromorphic Hardware", *PMES Workshop*.

T. Albash, I. Hen, F. M. Spedalieri, D. A. Lidar (2015) "Reexamination of the evidence for entanglement in the D-Wave processor," *Phys. Rev. A 92*, 62328, arXiv:1506.03539v2.

T. Albash, D. A. Lidar (2015) "Decoherence in adiabatic quantum computation," *Phys. Rev. A 91*, 062320, arXiv:1503.08767v2.

B. Alidaee, F. Glover, G. Kochenberger and C. Rego (2005) " A New Modeling and Solution Approach for the Number Partitioning Problem", *Journal of Applied Mathematics and Decision Sciences* 9 (2), pp. 135-145.

B. Alidaee, G. Kochenberger, K. Lewis, M. Lewis, H, Wang (2008) *"A New Approach for Modeling and Solving Set Packing Problems," European Journal of Operational Research*, 186(2):504-512.

D. Aloise, S. Cafieri, G. Caporossi, P. Hansen, S. Perron, L. Liberti (2010) "Column generation algorithms for exact modularity maximization in networks", *Physical Review E*, 82(4), 046112.

Alpha-QUBO (2019) http://meta-analytics.net/Home/AlphaQUBO

M. H. S. Amin, C. J. S. Truncik, D. V. Averin (2008) "Role of Single Qubit Decoherence Time in Adiabatic Quantum Computation**,"** *Phys. Rev. A 80*, 022303, arXiv:0803.1196v2**.**

M. Anthony, E. Boros, Y. Crama, A. Gruber (2017) *"Quadratic Reformulations of Nonlinear Binary Optimization Problems," Mathematical Programming*, 162(1-2):115-144.

M. Z. Alom, B. Van Essen, A. T. Moody, D. P. Widemann, T. M. Taha (2017) "Quadratic Unconstrained Binary Optimization (QUBO) on neuromorphic computing system," *IEEE 2017 International Joint Conference on Neural Networks (IJCNN)*, doi 10.1109/ijcnn.2017.7966350.

M. Aramon, G. Rosenberger, E. Valiante, T. Miyazawa, H. Tamura and H. G. Katzgraber (2019) "Physics-Inspired Optimization for Quadratic Unconstrained Problems Using a Digital Annealer", *Frontiers in Physics*, Volumn 7, Pages 48.

J. J. Berwald, J. M. Gottlieb, E. Munch (2018) "Computing Wasserstein Distance for Persistence Diagrams on a Quantum Computer", *arXiv:1809.06433*

S. Boixo, T. F. Rønnow, S. V. Isakov, Z. Wang, D. Wecker, D. A. Lidar, J. M. Martinis, M. Troyer (2014*)* "Evidence for quantum annealing with more than one hundred qubits," *Nature Physics*, vol. 10, pp. 218-224.

M. Booth, S. P. Reinhardt, A. Roy (2017) "Partitioning Optimization Problems for Hybrid Classical/Quantum Execution," *D-Wave Technical Report Series 14-1006A-A*, github.com/dwavesystems/qbsolv/qbsolv_techReport.pdf.

E. Boros, P. Hammer (1991) "The Max-cut Problem and Quadratic 0-1 Optimization: Polyhedral Aspects, Relaxations and Bounds," *Annals of Operations Research* 33(3):151-180.

E. Boros, P. Hammer (2002) "Pseudo-Boolean Optimization," *Discrete Applied Mathematics*, 123(1):155-225.

E. Boros, P. Hammer, G. Tavares (2007) "Local Search Heuristics for Quadratic Unconstrained Binary Optimization (QUBO)," *Journal of Heuristics*, 13(2):99-132.

E. Boros, P. Hammer, X. Sun (1989) "The DDT method for quadratic 0-1 minimization," *RUTCOR Research Center*, RRR: 39-89.

E. Boros, P. L.Hammer,, R. Sun,, G. Tavares (2008) "A max-flow approach to improved lower bounds for quadratic unconstrained binary optimization (QUBO)," *Discrete Optimization*, Volume 5, Issue 2, pp. 501-529.

G. Chapuis, H. Djidjev, G. Hahn, G. Rizk (2018) "Finding Maximum Cliques on the D-Wave Quantum Annealer," To be published in: *Journal of Signal Processing Systems*, DOI 10.1007/s11265-018-1357-8.

J. Clark, T. West, J. Zammit, X. Guo, L. Mason, D. Russell (2019) "Towards Real Time Multi-robot Routing using Quantum Computing Technologies", *HPC Asia 2019 Proceedings of the International Conference on High Performance Computing in Asia-Pacific Region*, pages 111-119.

P. Date, R. Patton, C. Schuman, T. Potok (2019) "Efficiently embedding QUBO problems on adiabatic quantum computers", *Quantum Information Processing 2019*, 18:117 https://doi.org/10.1007/s11128-019-2236-3.

E. P. Debenedictis (2019) "A Future with Quantum Machine Learning," *IEEE Computing Edge*, Vol. 5, No. 3, pp. 24 – 27.

N. Elsokkary, F.S. Khan, T. S. Humble, D. L. Torre, J. Gottlieb (2017) " Financial Portfolio Management using D-Wave's Quantum Optimizer: The Case of Abu Dhabi Securities Exchange", *2017 IEEE High-performance Extreme Computing (HPEC*).

E. Farhi and J. Goldstone (2014) "A Quantum Approximate Optimization Algorithm", *arXiv:1411.4028.*

S. Feld, C. Roch, T. Gabor, C. Seidel, F. Neukart, I. Galter, W. Mauerer, C. Linnhoff-Popien (2018) "A Hybrid Solution Method for the Capacitated Vehicle Routing Problem Using a Quantum Annealer", *arXiv:1811.07403*

F. Glover (1977) *"*Heuristics for Integer Programming Using Surrogate Constraints*," Decision Sciences*, Vol. 8, No. 1, pp. 156-166.

F. Glover (1996) "Tabu Search and Adaptive Memory Programming - Advances, Applications and Challenges," in *Interfaces in Computer Science and Operations Research*, Barr, Helgason and Kennington (eds.) Kluwer Academic Publishers, Springer, pp. 1-75.

F. Glover (1997) "A Template for Scatter Search and Path Relinking," in *Artificial Evolution, Lecture Notes in Computer Science*, 1363, J.-K. Hao, E. Lutton, E. Ronald, M. Schoenauer and D. Snyers, Eds. Springer, pp. 13-54.

F. Glover and G. Kochenberger, eds. (2003) *Handbook of Metaheuristics (International Series in Operations Research & Management Science)* Volume 1, Kluwer Academic Publishers, Springer, Boston.

F. Glover and G. Kochenberger (2019) "Quantum Bridge Analytics & QUBO 2.0," *Quantum Insight Conference 2019*, invited presentation 10/04/19, LHOFT – Luxembourg House of Financial Technology, 9, rue du Laboratoire, Luxembourg.

F. Glover and M. Laguna (1997) *Tabu Search*, Kluwer Academic Publishers, Springer.
F. Glover, B. Alidaee, C. Rego, G. Kochenberger (2002) "One-Pass Heuristics for Large Scale Unconstrained Binary Quadratic Problems," *European Journal of Operational Research*, 137(2):272-287.

F. Glover, G. Kochenberger, B. Alidaee (1998) "Adaptive Memory Tabu Search for Binary Quadratic Programs," *Management Science*, 44(3):336-345.

F. Glover, G. Kochenberger, B. Alidaee, M. Amini (1999) "Tabu Search with Critical Event Memory: An Enhanced Application for Binary Quadratic Programs," In: *Mete-Heuristics*, Springer, Berlin, pp. 93-109.

F. Glover, G. Kochenberger, B. Alidaee, M. Amini (2002*)* "Solving Quadratic Knapsack Problems by Reformulation and Tabu Search," *Combinatorial and Global Optimization*: (eds. P.M. Pardalos, A. Megados, R. Burkard, World Scientific Publishing Co., pp. 272-287.

F. Glover, G. Kochenberger, Y Wang (2018) "A new QUBO model for unsupervised machine learning," Research in progress.

F. Glover, J. Mulvey, D. Bai, and M. Tapia (1998) "Integrative Population Analysis for Better Solutions to Large-Scale Mathematical Programs," in *Industrial Applications of Combinatorial Optimization*, G. Yu, Ed.  Kluwer Academic Publishers, Springer, Boston, MA, pp. 212-237.

F. Glover, M. Lewis, G. Kochenberger (2018*)* "Logical and Inequality Implications for Reducing the Size and Difficulty of Unconstrained Binary Optimization Problems," *European Journal of Operational Research*, 265(2018) 829-842

F. Glover, Y. Tao, A. Punnen, G. Kochenberger (2015) "Integrating Tabu Search and VLSN Search to Develop Enhanced Algorithms: A Case Study Using Bipartite Boolean Quadratic Programs," *European journal of Operational Research*, 241(20;697-707.

G. Hahn and H. Djidjev (2017) "Reducing Binary Quadratic Forms for More Scalable Quantum Annealing", *2017 IEEE International Conference on Rebooting Computing*, DOI**:** 10.1109/ICRC.2017.8123654.

K. Hamilton, C.D. Schuman, S. R. Young, N. Imam and T. S. Humble (2018) "Neural Networks and Graph Alogrithms with Next-Generation Processors", *2018 IEEE International Parallel and Distributed Processing Symposium Workshops (IPDPSW)*. DOI: 10.1109/IPDPSW.2018.00184

P. Hammer, P. Hansen, B. Simeone (1984) "Roof Duality, Complementation and Persistency in Quadratic 0-1 Optimization," *Mathematical Programming*, 28(2):121-155.

H. H. Hoos (2012) "Programming by Optimization," *Communications of the ACM*, Vol. 55, Issue 2, pp. 70-80.

H. Huang, P. Pardalos, O. Prokepyev (2006) "Lower Bound Improvement and Forcing Rule for Quadratic Binary Programming," *Comput Optim Applied*, 33(2-3):187-208.

A. Kalra, F. Qureshi, M. Tisi (2018) "Portfolio Asset Identification using Graph Algorithms on a Quantum Annealer", http://www.henryyuen.net/fall2018/projects/qfinance.pdf

G. Kochenberger, A. Badgett, R. Chawla, F. Glover, Y. Wang and Y. Du (2019) "Comparison of QAOA and Alpha QUBO algorithms", in process.

G. Kochenberger and F. Glover (2006) "A Unified Framework for Modeling and Solving Combinatorial Optimization Problems: A Tutorial," In: *Multiscale Optimization Methods and Applications*, eds. W. Hager, S-J Huang, P. Pardalos, and O. Prokopyev, Springer, pp. 101-124.

G. Kochenberger, B. Alidaee, F. Glover, H. Wang (2007) "An Effective Modeling and Solution Approach for the Generalized Independent Set Problem," *Optimization Letters*, 1(1):111-117.

G. Kochenberger, F. Glover, B. Alidaee, C. Rego (2005*)* "An Unconstrained Quadratic binary Programming Approach to the Vertex Coloring Problem," *Annals of OR*, 139(1-4):229-241.

G. Kochenberger, F. Glover, B. Alidaee, H. Wang (2005) "Clustering of Micro Array Data via Clique Partitioning," *Journal of Combinatorial Optimization*, 10(1):77-92.

G. Kochenberger, F. Glover, B. Alidaee, K. Lewis (2005) "Using the Unconstrained Quadratic Program to Model and Solve Max 2-Sat Problems," *International Journal of OR*, 1(1):89-100.
G. Kochenberger, J-K Hao, S. Lu, H. Wang, F. Glover (2013) "Solving Large Scale Max Cut Problems via Tabu Search," *Journal of Heuristics*, 19(4):565-571.

G. Kochenberger, J-K. Hao, F. Glover, M. Lewis, Z. Lu, H. Wang, Y. Wang (2014) "The Unconstrained Binary Quadratic Programming Problem: A Survey," *Journal of Combinatorial Optimization*, Vol. 28, Issue 1, pp. 58-81.

G. Kochenberger and M. Ma (2019) "Quantum Computing Applications of QUBO Models to Portfolio Optimization," White paper, University of Colorado, Denver, September, 2019.

Lanting, A.J. Przybysz, A. Yu. Smirnov, F.M. Spedalieri, M.H. Amin, A.J. Berkley, R. Harris, F. Altomare, S. Boixo, P. Bunyk, N. Dickson, C. Enderud, J.P. Hilton, E. Hoskinson, M.W.

Johnson, E. Ladizinsky, N. Ladizinsky, R. Neufeld, T. Oh, I. Perminov, C. Rich, M.C. Thom, E. Tolkacheva, S. Uchaikin, A.B. Wilson, G. Rose (2014) "Entanglement in a quantum annealing processor," *Phys. Rev. X 4*, 021041, arXiv:1401.3500v1.

M. Lewis, B. Alidaee, F. Glover, G. Kochenberger (2009) "A Note on xQx as a Modeling and Solution Framework for the Linear Ordering Problem," *International Journal of OR*, 5(2):152-162.

M. Lewis, B. Alidaee, G. Kochenberger (2005) "Using xQx to Model and Solve the Uncapacitated Task Allocation Problem," *Operations Research Letters*, 33(2):176-182.

M. Lewis, G. Kochenberger, B. Alidaee (2008) "A New Modeling and Solution Approach for the Set Partitioning Problem." *Computers and OR*, 35(3):807-813.

A. Lucas (2014) "Ising Formulations of Many NP Problems," *Frontiers in Physics*, vol. 5, no. arXiv:1302.5843, p. 2.

Meta-Analytics(2019) http://meta-analytics.net/

A. Martin, E. Boros, Y. Crama, A. Gruber (2016) "Quadratization of Symmetric Pseudo-Boolean Functions," *Discrete Applied Mathematics* 203:1-12.

S. Mniszewski, C. Negre, H. Ushijima-Mwesigwa (2016). "Graph Partitioning using the D-Wave for Electronic Structure Problems," *LA-UR-16-27873*, 1–21.

S. M. Mniszewski, C. F. A. Negre, Ushijima-Mwesigwa (2018) "Graph Clustering Approaches using Nearterm Quantum Computing", *Argonne Quantum Computing Workshop.*

C. F. A. Negre, H. Ushijima-Mwesigwa, and S. M. Mniszewsk (2019) "Detecting Multiple Communities Using Quantum Annealing on the D-Wave System". *arXiv:1901.09756*

F. Neukart, G. Compostella, C. Seidel, D. Dollen, S. Yarkoni, B. Parney (2017) "Traffic flow optimization using a quantum annealer", *arXiv:1708.01625*

M. Ohzeki, A. Miki, M.J. Miyama, M.Terabe (2018) "Control of automated guided vehicles without collision by quantum annealer and digital devices", *arXiv:1812.01532*

D. O'Malley, V.V. Vesselinov, B. S. Alexandrov, L.B. Alexandrov (2018) "Nonnegative/Binary matrix factorization with a D-Wave quantum annealer". *PLoS ONE 13(12): e0206653.* https://doi.org/10.1371/journal.pone.0206653

Kerberos (2019) *Kerberos: The Network Authentication Protocol,* https://web.mit.edu/kerberos/.

S. Pakin (2017) "Navigating a Maze using a Quantum Annealer", *Proceedings of the Second International Workshop on Post Moores Era Supercomputing*, Pages 30-36.

S. Pakin (2018*)* "QMASM—Quantum macro assembler," https://ccsweb.lanl.gov/~pakin/software/ and https://github.com/lanl/qmasm

G. Palubeckis (2006) "Iterated Tabu Search for the Unconstrained Binary Quadratic Optimization Problem," *Informatica*, 17(2): 279-296

P. Pardalos, G. Rodgers (1990) "Computational Aspects of a Branch and Bound Algorithm for Quadratic zero-one Programming." *Computing*, 45(2):131-144.

P. Pardalos, J. Xue (1999) "The Maximum Clique Problem," *Journal of Global Optimization*, 4(3):301-328.

P. Pardalos, O. Prokopyev, O. Shylo, V. Shylo (2008) "Global Equilibrium Search Applied to the Unconstrained Binary Quadratic Optimization Problem," *Optimization Methods and Software*, 23(1):129-140.

E. Pelofske, G. Hahn, and H. Djidjev (2019) "Solving large Maximum Clique problems on a quantum annealer", *arXiv:1901.07657*

J. Preskill (2019) "Why I Called it 'Quantum Supremacy'," *Quanta Magazine*, https://www.quantamagazine.org/john-preskill-explains-quantum-supremacy-20191002/

K. L. Pudenz and D.A. Lidar (2013). "Quantum adiabatic machine learning," *Quantum information processing*, 12(5), 2027-2070.

Qbsolv (2017). D-Wave Initiates Open Quantum Software Environment. www.dwavesys.com/press-releases/d-wave-initiates-open-quantum-software-environment.

C. Reedy (2017) "When will quantum computers be consumer products?" *Futurism*, https://futurism.com/when-will-quantum-computers-be-consumer-products

S. Reinhardt (2018) "Detecting Lateral Movement with a Compute-Intense Graph Kernel", http://www.clsac.org/uploads/5/0/6/3/50633811/reinhardt-clsac-2018.pdf

E. Rodriguez-Heck (2018) "Linear ad Quadratic Reformulations of Nonlinear Optimization Problems in Binary Variables," *PhD Dissertation, Liege University*

I. Rosenberg (1975) "Reduction of Bivalent Maximization to the Quadratic Case,' Cahiers du Centre d'Etudes de Recherche Operationnelle," 17:71-74.

D. Sahner (2018) "A Potential Role for Quantum Annealing in the Enhancement of Patient Outcomes?", https://www.dwavesys.com/sites/default/files/Sahner.2018.pdf

M. Samorani, Y. Wang, Y. Wang, Z. Lu, F. Glover (2018) "Clustering-Driven Evolutionary Algorithms: An Application of Path Relinking to the Quadratic Unconstrained Binary Optimization Problem," to appear in the *Special Issue on Learning, Intensification and Diversification of the Journal of Heuristics*.

E. Schneidman, M. J. Berry, R. Segev; W. Bialek (2006), "Weak pairwise correlations imply strongly correlated network states in a neural population," *Nature*, 440 (7087): pp. 1007–1012.

R. Shaydulin, H. Ushijima-Mwesigwa, I. Safro, S. Mniszewski, Y. Alexeev (2018) "Community Detection Across Emerging Quantum Architectures," *PMES Workshop*.

V. Shylo, O. Shylo (2011) "Systems Analysis Solving Unconstrained Binary Quadratic Programming Problems by Global Equilibrium Search," *Cybern Syst Anal*, 47(6):889-897.

T. Simonite (2018a) "The Wired Guide to Quantum Computing," *Business* 8.24.18, https://www.wired.com/story/wired-guide-to-quantum-computing/.

T. Simonite (2018b) "It's Time You Learned About Quantum Computing," *Business*, 6.25.28, www.wired.com/story/time-you-learned-about-quantum-computing/

The National Academies of Sciences, Engineering and Medicine Consensus Study Report (2019), *Quantum Computing: Progress and Prospects* https://www.nap.edu/catalog/25196/quantum-computing-progress-and-prospects.

H. Ushijima-Mwesigwa, C. F. A. Negre, and S. M. Mniszewsk (2017) "Graph Partitioning using Quantum Annealing on the D-Wave System", *arXiv:1705.03082*.

A. Verma, M. Lewis (2019) "Optimal quadratic reformulations of fourth degree Pseudo-Boolean functions", *Optimization Letters*, https://doi.org/10.1007/s11590-019-01460-7

T. Vyskocil, S. Pakin, and H. N. Djidjev (2019) " Embedding Inequality Constraints for Quantum Annealling Optimization," *Quantum Technology and Optimization Problems. QTOP 2019. Lecture Notes in Computer Science,* vol 11413. Springer, Cham

H. Wang, B. Alidaee, F. Glover, G. Kochenberger (2006) "Solving Group Technology Problems via Clique Partitioning," *International Journal of Flexible Manufacturing Systems*, 18(2):77-87.

H Wang, Y. Wang, M. Resende, and G. Kochenberger (2016) "A QUBO Approach to Solving QAP Problems," Unpublished manuscript.

Q. Wang and T. Abdullah (2018) "An Introduction to Quantum Optimization Approximation Algorithm," https://www.cs.umd.edu/class/fall2018/cmsc657/projects/group_16.pdf

Y. Wang, Z. Lu, F. Glover and J-K. Hao (2012) "Path relinking for unconstrained binary quadratic programming," *European Journal of Operational Research* 223(3): pp. 595-604.

Y. Wang, Z. Lu, F. Glover and J-K. Hao (2013) "Backbone guided tabu search for solving the UBQP problem," *Journal of Heuristics*, 19(4): 679-695.

Z. Wang, S. Hadfield, Z. Jiang, and E. G. Rieffel (2017) "The Quantum Approximation Optimization Algorithm for MaxCut: A Fermionic View," *arXiv:1706.02998*.

H. Yu, Y. Huang and B. Wu (2018) "Exact Equivalence between Quantum Adiabatic Algorithm and Quantum Circuit Algorithm," *arXiv: 1706.07646v3* [quant-ph], DOI: 10.1088/0256-307X/35/11/110303.

L. Zhou, S. Wang, S. Choi, H.Pichler, and M. D. Lukin (2018) "Quantum Approximate Optimization Algorithm: Performance, Mechanism, and Implementation on Near-Term Devices",  *arXiv:1812.01041*

## Acknowledgements:

This tutorial was influenced by our collaborations on many papers over recent years with several colleagues to whom we owe a major debt of gratitude. These co-workers, listed in alphabetical order, are: Bahram Alidaee, Dick Barr, Andy Badgett, Rajesh Chawla, Yu Du, Jin-Kao Hao, Mark Lewis, Karen Lewis, Zhipeng Lu, Abraham Punnen, Cesar Rego, Yang Wang, Haibo Wang and Qinghua Wu.  Other collaborators whose work has inspired us are too numerous to mention.  Their names may be found listed as our coauthors on our home pages.